\documentclass[useAMS,usenatbib]{mn2e}
\pdfoutput=1
\usepackage{graphicx}
\usepackage{subfigure}
\usepackage[english]{babel}
\usepackage{amssymb}
\usepackage{xspace}
\usepackage{hyperref}

\newcommand\ionm[2]{#1$\,${\small\rmfamily{#2}}} 
\newcommand{\hi}{\ionm{H}{I}\xspace}
\def\hii{\ifmmode {\mbox H{\scshape ii}}\else H{\scshape ii}\fi\xspace}
\def\h2{\ifmmode {\mbox H$_2$}\else H$_2$\fi\xspace}
\def\frach2star{$\frac{\rm{M}_{\rm H2}}{\rm{M}_{\star}}$}

\title [Gas evolution of galaxies in dark matter haloes]{Evolution of
  the atomic and molecular gas content of galaxies in dark matter haloes}
\author[G. Popping, P.S. Behroozi and M.S. Peeples]{Gerg\"o Popping$^{1,2}$\thanks{E-mail: gpopping@eso.org},
  Peter S. Behroozi$^{3}$ and Molly S. Peeples$^{3}$\\
$^{1}$Kapteyn Astronomical Institute, University of Groningen, Postbus 800, NL-9700 AV Groningen, the Netherlands,\\
$^{2}$European Southern Observatory, Karl-Schwarzschild-Strasse 2,
85748, Garching, Germany\\
$^{3}$Space Telescope Science Institute, Baltimore, MD 21218 USA}
\begin{document}

\maketitle

\begin{abstract}
We present a semi-empirical model to infer the atomic and molecular
hydrogen content of galaxies as a function of halo mass and time. Our
model combines the SFR--halo mass--redshift relation (constrained
by galaxy abundances) with inverted SFR--surface density relations to infer galaxy \hi and \h2 masses. 
We present gas scaling relations, gas fractions, and mass
functions from $z=0$ to $z=3$ and the gas properties of
galaxies as a function of their host halo masses. Predictions of our
work include: 1) there is a $\sim 0.2$ dex decrease in the \hi mass of
galaxies as a function of their stellar mass since $z=1.5$, whereas
the \h2 mass of galaxies decreases by
$>1$ dex over the same period. 2) galaxy cold gas fractions and
\h2 fractions decrease with increasing stellar mass and time. Galaxies
with $M_{\star} > 10^{10}\,\rm{M}_\odot$ are dominated by their
stellar content at $z\leq$ 1, whereas less-massive galaxies
only reach these gas fractions at $z=0$. We find the strongest evolution
in relative gas content at $z<1.5$. 
3) the SFR to gas mass ratio decreases by an order
of magnitude from $z=3$ to $z=0$. This is accompanied by lower \h2 fractions; these lower fractions in combination with smaller gas reservoirs correspond to decreased present-day galaxy SFRs. 4) an \h2-based star-formation
relation can simultaneously fuel the evolution of the cosmic
star-formation and reproduce the observed weak evolution in the cosmic \hi
density. 5) galaxies residing in haloes with masses near $10^{12}\,\rm{M}_\odot$ are most
efficient at obtaining large gas reservoirs \emph{and} forming
\h2 at all redshifts. These two effects lie at the origin of the high star-formation
efficiencies in haloes with the same mass. 
\end{abstract}


\section{Introduction}
Understanding of the gas evolution of galaxies provides fundamental
constraints on our theories for galaxy formation and evolution. The
star formation (SF) in a galaxy is regulated by the amount of cold
gas eligible to form stars. Observations have shown a strong correlation
between star-formation rate (SFR) surface density and the density of
molecular hydrogen out to $z\sim 2$
\citep{Wong2002,Bigiel2008,Daddi2010,Genzel2010,Bigiel2012}. Locally, 
the correlation of SFR surface density with atomic hydrogen is weak
\citep{Bigiel2008,Leroy2008}. These results suggest that before
  forming stars, cool gas that accretes onto a galaxy must go through an
  atomic and molecular phase. Knowledge
of the cold gas content of galaxies and its partitioning into \hi and \h2 therefore provides
valuable constraints on the physics that control this baryonic cycle.

Abundance matching techniques and empirical models have shown that haloes with masses near 
$10^{12}\,\rm{M}_\odot$ are most efficient at forming stars and
that the star-formation histories for galaxies with varying $z=0$ halo
mass are very different
\citep[e.g.,][]{Yang2003,Conroy2009,Behroozi2010,Guo2010,Moster2010,Wang2010,Yang2012,Behroozi2013,Moster2013,Reddick2013,Luz2014,Luz2014giant_to_dwarf}. 
The halo mass dictates the gas supply onto a
galaxy. It is therefore crucial to couple the cold gas content of a galaxy to the halo
framework, in order to fully understand the physics that
translates the supply of cold gas into the
stellar buildup of galaxies and constrain processes such as winds, heating, and cooling.

Large surveys of the cold gas content in galaxies have mainly been restricted to our local Universe. Atomic hydrogen surveys such as ALFALFA
\citep{Giovanelli2005}, THINGS
\citep[The HI Nearby Galaxy Survey;][]{Walter2008}, GASS \citep[GALEX Arecibo SDSS
survey;][]{Catinella2010,Catinella2012,Catinella2013} and molecular
hydrogen surveys (typically through CO emission) such as BIMA SONG
\citep[Bima survey of nearby galaxies;][]{Helfer2003}, HERACLES
\citep[HERA CO-Line Extragalactic Survey;][]{Leroy2009_HERACLES}, and COLD GASS
\citep[CO legacy database for GASS;][]{Saintonge2011} have mapped
the \hi and \h2 content of local galaxies, providing constraints on the
gas-to-star ratios and \hi and \h2 mass function of galaxies. These
surveys have provided important insights into the connections between stellar mass, star formation rate and gas properties of galaxies,
but do not probe the evolution in galaxy gas properties that drive the stellar
buildup of galaxies. 

Observations of atomic gas in emission in galaxies have been
restricted to $z\leq 0.25$
\citep{Verheijen2007,Catinella2008,Catinella2015}. As a pilot for CHILES
  (Cosmos \hi Legacy Survey), \citet{Fernandez2013} observed
\hi in emission in 33 galaxies in the redshift interval $0<z<0.19$. Damped Lyman-$\alpha$ absorbers (DLAs) have provided an alternative approach to estimate the global
gas content in galaxies up to much higher redshifts
\citep[$z<4$; e.g.,][]{Rao2006,Prochaska2009,Noterdaeme2012,Zafar2013},
but the nature of DLAs and their connection to galaxies is still
unclear \citep{Berry2013}.

Over the past years a wealth of direct observations of \h2 (through CO) in
high-redshift galaxies have become available, although usually biased
towards gas-rich, actively star-forming galaxies
\citep[e.g.,][]{Daddi2010,Tacconi2010,Genzel2010,Geach2011,Riechers2011,Bauermeister2013,Tacconi2013}. Although results from these surveys are still inconclusive because of small and biased samples, these studies suggest that galaxies at high-redshift may
have had larger molecular hydrogen reservoirs than local
counterparts. The above efforts have led to valuable insights and constraints on
galaxy formation models, but still show a very limited picture of the
evolution of cold gas in galaxies. We anticipate that the newest
generation of radio and sub-mm instruments such as ALMA, SKA and its
pathfinders MeerKat and ASKAP will improve the number statistics and
revolutionize our understanding of the gas properties of galaxies over
cosmic time.

Indirect estimates of the total gas and/or \h2 mass of large samples
of galaxies at high redshift \citep[Popping et al. in prep.]{Erb2006,Mannucci2009,Popping2012} have improved the
number statistics but still suffer from severe selection
criteria. These studies were not in the position to address
the \hi and \h2 mass function, nor the cosmic density of cold
gas. Furthermore, they lack the capability to probe the connection
between a galaxy's halo and cold gas content.

In the last decade significant progress has been made on developing models that
track the \hi and \h2 content of galaxies
\citep[e.g.,][]{Obreschkow2009_sam,Dutton2010,Fu2010,Lagos2011cosmic_evol,Christensen2012,Krumholz2011,Kuhlen2012,Dave2013,Popping2013,Rafieferantsoa2014,Thompson2014}. These models have proven to be very successful in reproducing the available
observational constraints on the \hi, \h2 and sub-mm
\citep{Popping2013RT} properties of galaxies in
the local and high-redshift Universe. Despite this success, theoretical
models still have a hard time reproducing stellar-to-halo mass ratios
at $z>0$ \citep{Lu2013} and the star-formation history
of low-mass galaxies \citep{Weinmann2012,Somerville2014}. Additional information on
the gas content of galaxies in haloes will be crucial to break
degeneracies in different physical mechanisms included in models such
as supernovae feedback, metal enrichment, feedback from active 
galactic nuclei and shock heating. 

In this paper we present a new semi-empirical approach to constrain the \hi and \h2
content of galaxies as a function of time and halo mass in the
redshift range $z=0 - 3$. We couple a sub-halo abundance
matching model (SHAM) \citep{Behroozi2013SFE,Behroozi2013} with a model to
indirectly estimate the \hi and \h2 content of a galaxy
\citep{Popping2012}. With our approach we can follow galaxy gas properties as a function of time and halo
mass. It provides an unique
opportunity to infer galaxy \hi and \h2 masses, mass functions, and cosmic densities
out to $z=3$. This allows us to gain insights into the processes that shape the maximum star-formation
efficiency in haloes with $M_{\rm vir} \sim 10^{12}\,\rm{M}_\odot$ and study the gas
properties of galaxies that drive the peak in cosmic SFR density at
$z\sim 2$ \citep[e.g.][]{Hopkins2006,Madau2014}. We aim to provide constraints for the gas content of
galaxies to be observed with the newest generation of radio and
sub-mm instruments and at the same time provide valuable
constraints for galaxy formation models. One of the main benefits in choosing this
approach is that galaxy gas masses are based on star-formation
histories that {\it by construction} are representative of real
galaxies. The presented approach is in stark contrast to semi-analytical
  models. Instead of having to
introduce uncertain recipes for the formation and evolution of
galaxies, we can focus on the galaxy gas content within a
realistic framework. Similar techniques have been used to look at the
evolution of the stellar metallicity and galaxy age
\citep{Peeples2013,Munoz2014} and the in- and outflow of gas in
galaxies \citep{LuMoLu2014}. Because our approach is based on
  realistic star-formation histories, the inferred gas properties
  provide very useful predictions for future observations.

This paper is organized as follows. In Section \ref{sec:methodology}
we present our methodology. In Section \ref{sec:results} we present
our results and we discuss them in Section \ref{sec:discussion}. We
summarize our findings in Section \ref{sec:summary}. Throughout this
paper we assume a flat $\Lambda$ cold dark matter ($\Lambda$CDM) cosmology
with $\rm{H}_0 = 70 \,\rm{km}\,{s}^{-1}$, $\Omega_{\rm matter} =
0.27$, $\Omega_\Lambda = 0.73$, $n_s = 0.95$, and $\sigma_8 = 0.82$ \citep{Behroozi2013}. We
assume a Chabrier stellar initial mass function
\citep[IMF:][]{Chabrier2003} in the mass range $0.1 -
100\,\rm{M}_\odot$ and where necessary convert observational
quantities used to a Chabrier IMF. All presented gas masses are
  pure hydrogen masses. Unless noted different, presented gas masses do not include a correction for
  Helium.

\begin{figure*}
\includegraphics[width = 0.65\hsize]{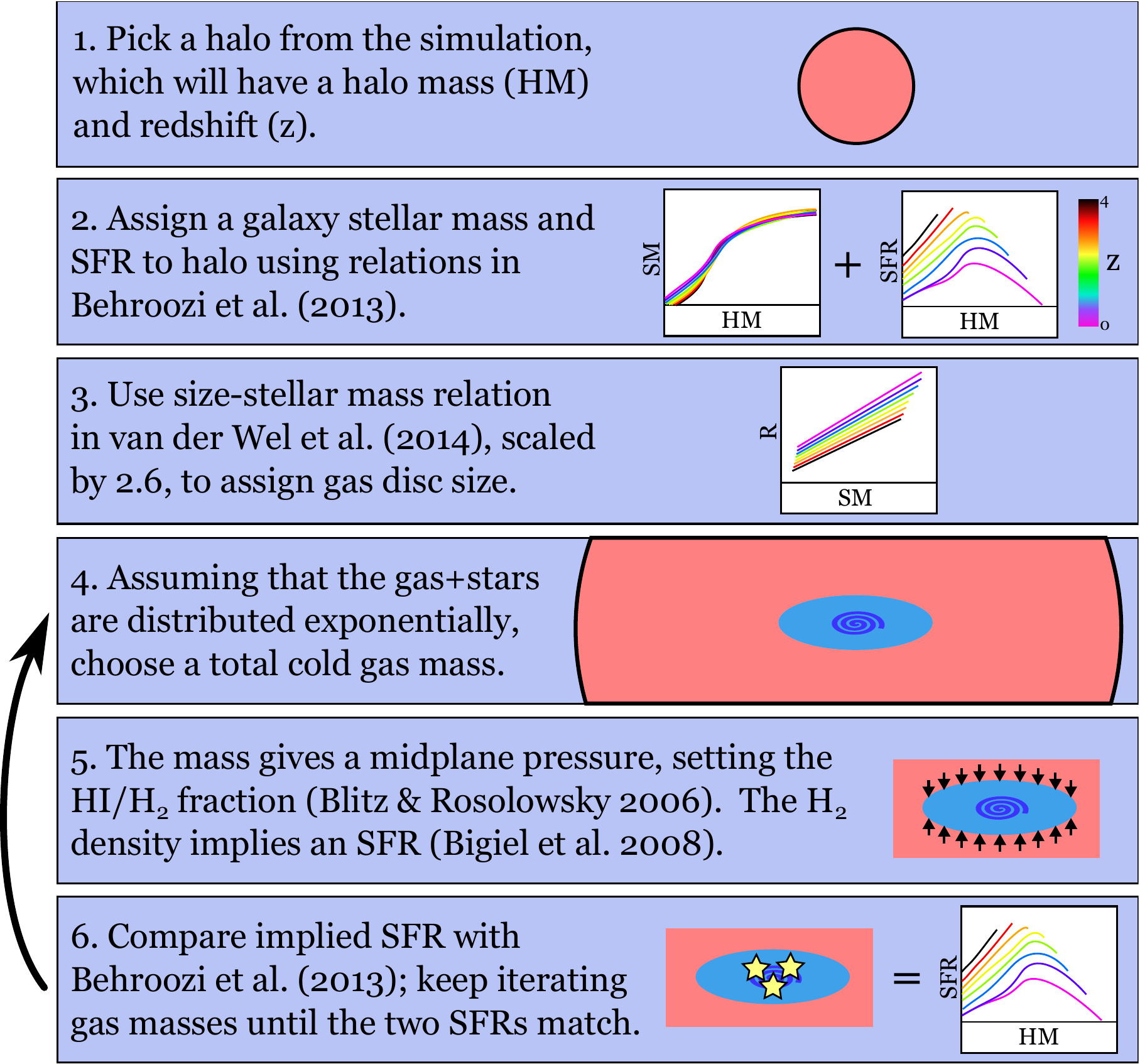}
\caption{A flowchart of the methodology used to estimate the \hi and
  \h2 masses of galaxies based on realistic star-formation histories.
\label{fig:flowchart}}
\end{figure*}

\section{Methodology}
\label{sec:methodology}
This section describes our methodology to model the gas content of
galaxies. We will first give a short overview of the SHAM used to
provide SFRs, stellar masses, and halo masses in Section \ref{sec:sham}. We will then describe
our method to estimate the gas masses of galaxies in Section
\ref{sec:model}. We present a schematic overview of this model and its
main individual ingredients in Figure \ref{fig:flowchart}. The
individual steps depicted in Figure \ref{fig:flowchart} are described below.

\subsection{Sub-halo abundance matching}
\label{sec:sham}
We use the abundance modeling approach in \citep{Behroozi2013} to provide
star-formation histories (halo mass, stellar mass, and SFR as a
function of redshift) that by construction are representative of
real galaxies. The abundance modeling approach adopts a very
flexible parametrization for the stellar mass--halo mass relationship
as a function of redshift, $M_\star(M_{\rm vir}, z)$.  This parametrization has six
variables to control the normalization and shape of the relationship
(a characteristic stellar mass, halo mass, faint-end slope,
massive-end cutoff, transition region shape, and scatter); for each
variable, three parameters control the redshift scaling at low
($z=0$), mid ($z\sim 1$) and high ($z>3$) redshift.  This results in a
total of eighteen shape parameters. Additional nuisance parameters
account for systematic observational biases in how galaxy luminosities
are converted to stellar masses/star formation rates.  A specific
choice of $M_\star(M_{\rm vir}, z)$ applied to a dark matter simulation results in
an assignment of a galaxy stellar mass to every halo at every redshift
in the simulation.  By counting galaxies at individual redshifts, and
by tracing the growth of galaxies along dark matter halo merger trees,
this assignment yields concrete predictions for observed stellar mass
functions, cosmic star formation rates, and specific star formation
rates.  Comparing these predictions to observed data (i.e., stellar
mass functions and star formation rates) from $z=0$ to $z\sim8$
results in a likelihood for a given choice of $M_\star(M_{\rm vir}, z)$.  The
posterior distribution for $M_\star(M_{\rm vir},z)$, along with derived values for
the average star formation rate as a function of halo mass and
redshift, are then inferred from observed data using an MCMC approach.
The full details of this method are presented in \citet{Behroozi2013}.

The average SFRs include a contribution from actively
star-forming and quenched galaxies. In this paper we are particularly interested in
star-forming galaxies.  To estimate the fraction of quenched galaxies we use the fitting
formula by \citet{Brammer2011}
\begin{equation}
  \label{eq:Brammer2011}
  f_{\rm quenched}(M_\star,z) = \bigl [\bigl ( \frac{M_\star}{10^{10.2 +
      0.5z}\,M_\odot}\bigr )^{-1.3} + 1 \bigr ] ^{-1}.
\end{equation}
Beyond $z=3$ all galaxies are considered to actively form stars
(i.e., $f_{\rm quenched} = 0$). Under the assumption that the
contribution from quenched galaxies to the average SFR is negligible
we can estimate the SFR of actively star-forming galaxies as $SFR_{\rm
  active}= SFR_{\rm average} \,/ \,(1 - f_{\rm quenched} )$. Combined
with the evolution of the SFR -- stellar mass relation, this quenched
fraction evolution accurately reproduces the distribution of ages of
local galaxies \citep{Munoz2014}.

To reproduce the observed scatter in the relation between stellar mass
and SFR we calculate the gas masses for each halo mass--stellar mass--SFR realization multiple times with a scatter in SFR of 0.3 dex
\citep[e.g.,][]{Noeske2007}. We assume a scatter in the stellar
  masses at fixed halo mass of  $\xi = 0.218 - 0.023 (a - 1)$ dex, where $a$ is
  the cosmological scale factor. On top of that we assume that the
  observational uncertainty around the
stellar masses is log-normal, with a redshift dependent standard
deviation $\sigma(z) = \sigma_0 + \sigma_zz$, where $\sigma_0 = 0.07$
and $\sigma_z = 0.061$ \citep{Behroozi2013}.

Halo mass functions used in this work are taken from the fitting
functions in \citet{Behroozi2013}, which are based on the fits in \citet{Tinker2008}.  Halo masses are defined as spherical overdensities
according to the virial overdensity criterion of \citet{Bryan1998}.

\subsection{Indirect gas measures}
\label{sec:model}
We infer the \hi and \h2 content of a galaxy based on the approach
presented in \citet{Popping2012}. This method uses the combination of an
empirical molecular-based SF relation \citep[based on][]{Bigiel2008} and a
pressure-based prescription to calculate the \h2-to-\hi
surface-density ratio of cold
gas \citep{Blitz2006}. We pick a gas mass and solve for the SFR using
the set of equations described below. We then iterate through gas
masses till convergence with the SFR picked from \citet{Behroozi2013}
is reached.

We use a slightly adapted version of the molecular-based SFR relation
deduced in \citet{Bigiel2008} to allow for higher star-formation
efficiencies in  high gas surface density regions. This is motivated by
the results of \citet{Daddi2010} and \citet{Genzel2010}, who found
that the SFR surface density relation at high densities has a slope of
1.4 (versus 1 in Bigiel et al. 2008). The adopted equation is
given by
\begin{equation}
\label{eq:bigiel}
\Sigma_{\mathrm{SFR}} = \frac{A_{\mathrm{SF}}}{10 {\rm M}_\odot\,\mathrm{pc}^{-2}}\,\left(1 + \frac{\Sigma_{\mathrm{gas}}}{\Sigma_{\mathrm{crit}}}\right)^{N_{\mathrm{SF}}}\,f_{\mathrm{H}_2}\,\Sigma_{\mathrm{gas}}
\end{equation}
where $\Sigma_{\mathrm{SFR}}$ and $\Sigma_{\mathrm{gas}}$ are the star
formation and cold gas surface densities in
$\rm{M}_\odot\,\mathrm{yr}^{-1}\,\mathrm{kpc}^{-2}$ and $\rm{M}_\odot\,\mathrm{pc}^{-2}$,
respectively; $A_{\mathrm{SF}}$ is the normalization of the power law in
$\rm{M}_\odot\,\mathrm{yr}^{-1}\,\mathrm{kpc}^{-2}$;
$\Sigma_{\mathrm{crit}}$ is the critical surface density above which the star formation follows
\citet{Kennicutt1998law}; $N_{\mathrm{SF}}$ is an index which sets the
efficiency; and
$f_{\mathrm{H}_2}=\Sigma_{\mathrm{H}_2}/(\Sigma_{\mathrm{HI}}+\Sigma_{\mathrm{H}_2})$
is the local molecular gas fraction. Following \citet{Popping2012} we use
$N_{\mathrm{SF}} = 0.5$ and $\Sigma_{\mathrm{crit}} = 100
\,\rm{M}_\odot\,\rm{pc}^{-1}$. When not accounting for the increased star-formation
  efficiency of high surface density regions (i.e. $N_{\mathrm{SF}} =
  0$) we infer 10 percent more gas in galaxies with $\rm{SFR}/(2\pi r_{\star}^2) \approx
  1\,\mathrm{M}_\odot\,yr^{-1}\,kpc^{-2}$ and roughly 75 percent more
  gas in galaxies with $\rm{SFR}/(2\pi r_{\star}^2) \approx
  10\,\mathrm{M}_\odot\,yr^{-1}\,kpc^{-2}$.

We calculate the molecular fraction of the cold gas using a pressure-regulated recipe to determine the
molecular fraction of the cold gas, based on the work by
\citet{Blitz2006}. The authors found a power-law relation between the
mid-plane pressure acting on a galaxy disc and the ratio between
molecular and atomic hydrogen, i.e.,
\begin{equation}
R_{\mathrm{H}_2} = \left(\frac{\Sigma_{\mathrm{H}_2}}{\Sigma_{\mathrm{HI}}}\right) = \left(\frac{P_m}{P_0}\right)^\alpha
\label{eq:blitz2006}
\end{equation}
where $P_0 $ is the external pressure in the interstellar medium where the
molecular fraction is unity and $\alpha$ is the power-law
index. We use $P_0 = 3.25\times10^{-13} \mathrm{erg\,cm^{-3}}$ and
$\alpha = 0.8$, based on the results presented in \citet{Leroy2008}. $P_m$ is the mid-plane pressure acting on the
galaxy disc. Following \citet{Popping2012} we
describe the mid-plane pressure by
\begin{equation}
P_m(r) = \frac{\pi}{2}\,G\,\Sigma_{\mathrm{gas}}(r)\left[\Sigma_{\mathrm{gas}}(r) + 0.1
\sqrt{\Sigma_{\star}(r)\Sigma_{\star,0}}\right],
\label{eq:pressure}
\end{equation}
where G is the gravitational constant, $r$ is the radius from the
galaxy centre, and $\Sigma_{\star, 0} \equiv M_{\star}/(2 \pi
r_{\star}^2)$, based on empirical scalings for nearby disc galaxies. We now have all the necessary ingredients to calculate
$R_{\mathrm{H}_2}$ and subsequently the cold gas molecular fraction
[$f_{\mathrm{H}_2}=R_{\mathrm{H}_2} /(1+R_{\mathrm{H}_2})]$. 

We assume that cold gas and stars are distributed following an
exponential profile\footnote{This is slightly different from
  \citet{Popping2012}, where the star formation was assumed to follow
  an exponential distribution rather than the cold gas.}. We estimate the
disc scale length $r_{\star}$ as a function of stellar mass and
redshift by linearly interpolating
through the fits presented in Table 2 in
\citet{vanderWel2014}. Stellar disc sizes increase with stellar mass
and time. For galaxies with stellar masses less than
$10^{9}\,\rm{M}_\odot$ we estimate the disc scale length based on work by
\citet{Kravtsov2013}, who relates the scale length of a disc to the size of its
host halo. The typical uncertainty given by \citet{vanderWel2014} is
0.2 dex. 
Following
\citet{Kravtsov2013} we assume that the scale length of the gaseous disc
is 2.6 times larger than the scale length of the stellar disc.  Varying $\chi_{\rm gas}$
results in negligible differences in inferred cold gas and \h2 mass as
long as $\chi_{\rm gas} > 1$. Decreasing $\chi_{\rm gas}$ to values
less than one leads to more significant differences in the inferred
gas masses. When adopting $\chi_{\rm gas} = 0.5$ we find that the
inferred gas and molecular masses are lowered by 0.25 and 0.1 dex,
respectively.  Observations of the sizes of the CO discs
of galaxies at $z\sim1.2$ and $z\sim2.2$ (supposedly tracing the molecular hydrogen)
do not support a scale length ratio of $\chi_{\rm gas} \ll 1.0$
\citep[e.g.,][]{Tacconi2010,Tacconi2013}. Resolved observations of CO
discs with ALMA will be able to better address the scale length ratio
between the stellar and molecular disc of galaxies at high redshifts.

The assumption of an exponential disc morphology may not be physical
for galaxies at $z=0$ with stellar masses below $\sim
10^9\,\rm{M}_\odot$ \citep{Kelvin2014}. At higher
redshifts it is unclear at which stellar mass the assumption of disc
structure may break down. As a consequence, assuming that the
distribution of gas in irregular galaxies is clumpy, the inferred \hi content of galaxies in this
mass regime may be too high. The inferred \hi disc mass for satellites
may also be overestimated if these galaxies undergo significant stripping.

We calibrated our method using direct observations of the \hi and/or
\h2 content of galaxies in the local and high-redshift
Universe from \citet{Leroy2008}, \citet{Daddi2010},
\citet{Tacconi2010}, and \citet{Tacconi2013}. Using
$\chi^2$-minimization, we find best
agreement between inferred and observed gas masses for this sample when adopting a
value of $A_{\mathrm{SF}}=9.6 \times 10^{-3}\,\rm{M}_\odot\,\mathrm{yr}^{-1}\,\mathrm{kpc}^{-2}$ for the
efficiency of forming stars out of molecular gas. We integrate the
disc out to a hydrogen column density of $N_{H}
= 10^{18}\,\rm{cm}^{-2}$ (the cold gas at these column densities is
atomic). The adopted value for  $A_{\mathrm{SF}}$ is within the errors on $A_{\mathrm{SF}}$ in \citet{Bigiel2008} and
\citet{Bigiel2012}.  The cut in hydrogen column density at $N_{H}
= 10^{18}\,\rm{cm}^{-2}$ affects the inferred \hi masses by a few
percent in galaxies with global SFR surface density (defined as
$\rm{SFR}/[2\pi r_{\star}^2]$) less than $10^{-4}$ with respect to
the approach adopted in \citet{Popping2012}. 

\citet{Popping2012} showed that this method is very successful in
reproducing the cold gas and \h2 surface densities of local galaxies
and the integrated \hi and/or \h2 mass of local and high-redshift
galaxies. We refer the reader to the Popping et al. paper for a more
detailed description of this method, including uncertainties in the
free parameters. The typical systematic uncertainty for the inferred gas masses (based
on uncertainties in $\chi_{\rm gas}$, the star-formation efficiency,
and the parameters that control the partitioning of cold gas in an
atomic and molecular component) is 0.25 dex.

\begin{figure*}
\includegraphics[width = \hsize]{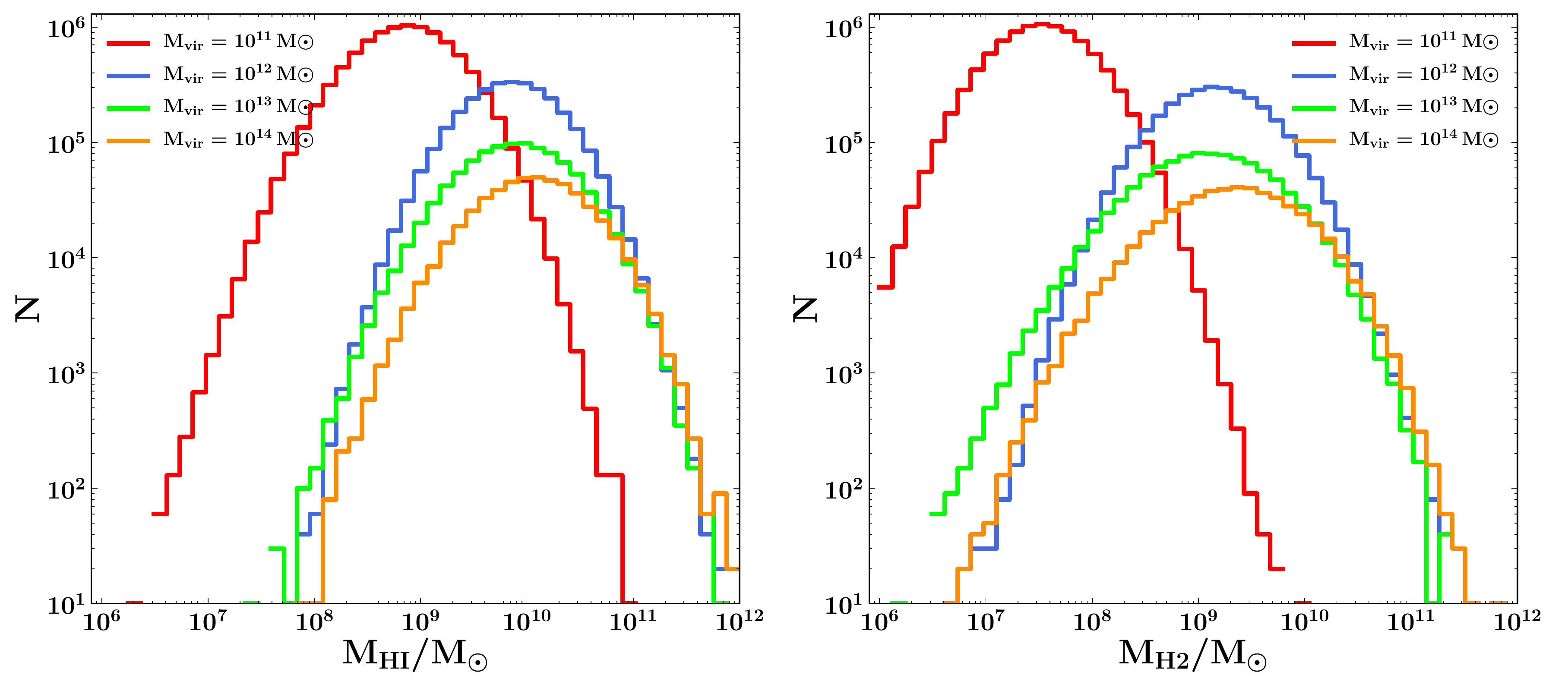}
\caption{Histograms of the distribution in the inferred \hi (left) and \h2
  (right) masses for fixed halo masses at $z=0.0$.
\label{fig:distribution}}
\end{figure*}

\subsection{Distribution of inferred gas masses for fixed halo mass}
This work is based on several relations with different underlying distributions and uncertainties. To develop a
feeling of the effect that these underlying distributions have on the
inferred gas masses, we show the histograms of the distributions of inferred \hi and
\h2 masses for fixed halo masses at $z=0.0$ in Figure \ref{fig:distribution}. We find that the distribution in \hi and \h2 masses is typically
log-normal. The distribution in \hi has a width of $\sim 0.45$ dex,
independent of host halo mass. The \h2 distribution has a width of
$\sim 0.45$ and $\sim 0.6$ dex for galaxies residing in haloes with
masses of $10^{11-12}\,\rm{M}_\odot$ and $10^{13-14}\,\rm{M}_\odot$,
respectively.

\begin{figure*}
\includegraphics[width = \hsize]{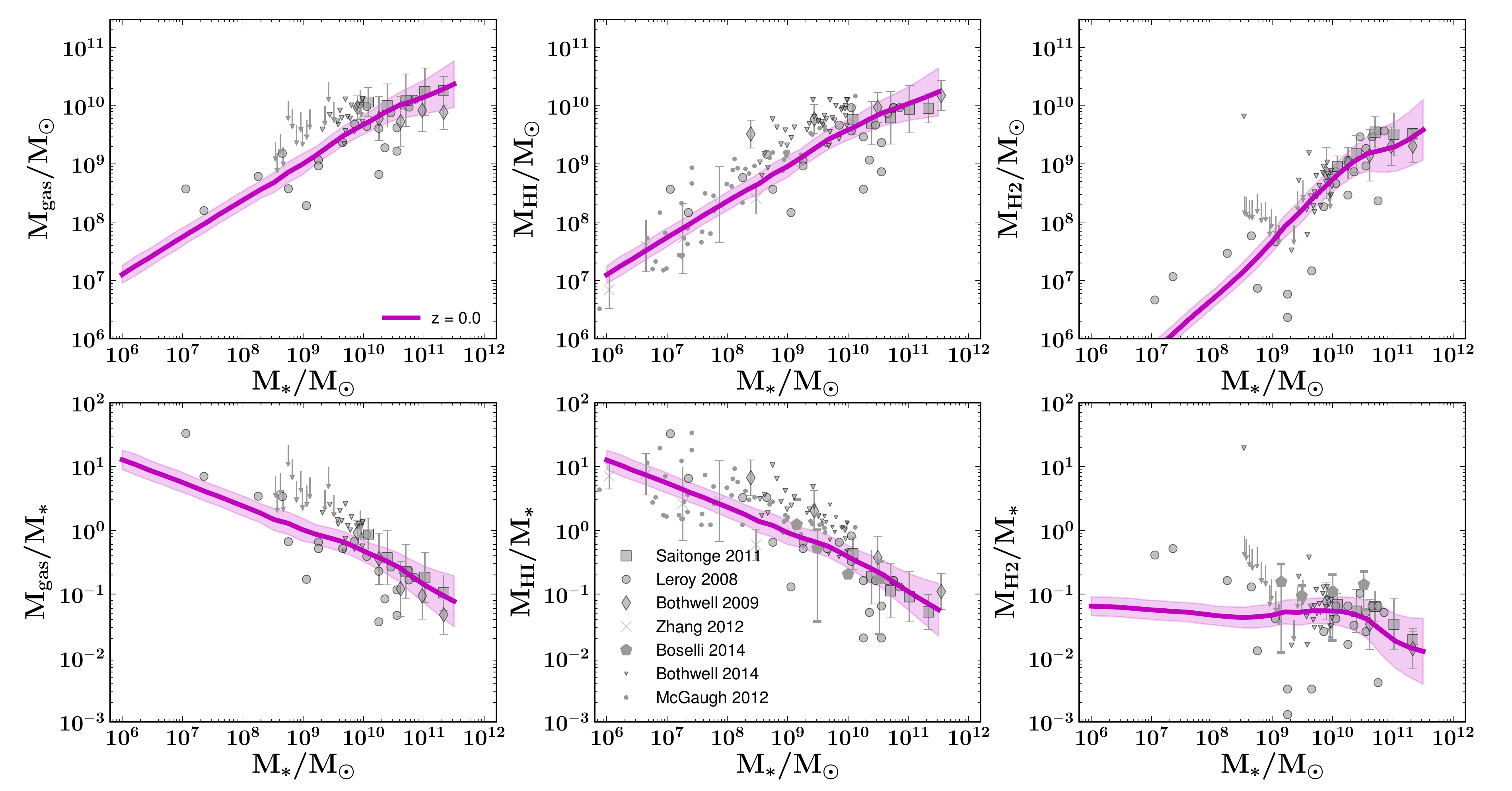}
\caption{The cold gas mass (top left), \hi mass (top row center
  column), \h2 mass (top right), gas-to-stellar mass ratio (bottom
  left), \hi-to-stellar mass ratio (bottom row center column), and \h2-to-stellar mass ratio
  (bottom right) of galaxies as
  a function of stellar mass at $z=0$. Shaded regions mark the one
  sigma of the posterior distribution of the inferred gas masses. Literature values are from
  \citet{Leroy2008}, \citet{Bothwell2009}, \citet{Saintonge2011},
  \citet{McGaugh2012}, \citet{Zhang2012}, \citet{Boselli2014} and \citet{Bothwell2014} are shown in
  grey. Errorbars on the literature values mark the one-sigma distributions
  of the observed sample.
\label{fig:scalez0}}
\end{figure*}

\section{Results}
\label{sec:results}
In this Section we present our inferred gas masses for galaxies in the local and high-redshift Universe. We compare our results to observations of local and
high-redshift galaxies in Section \ref{res:local}. We present
predictions for future \hi and \h2 surveys of galaxies at $z>0$  in Section \ref{res:highz}. We finish by discussing
the gas properties of galaxies as a function of their halo mass in
Section \ref{sec:gashalo}. In all figures the solid line and the
  shaded regions mark the mean and one standard deviation of the posterior
  distributions.  The posterior distributions are obtained by
  inferring the gas masses based on a thousand different realizations
  for the stellar mass -- halo mass relation. For each realization we
  infer the gas masses multiple times, to fully cover the
  uncertainties and distributions in SFR, stellar, mass, size, and our methodology to
  infer gas masses.

\subsection{Gas in local galaxies}
\label{res:local}
We present scaling relations for the total cold gas, \hi and \h2 content of
local star-forming galaxies compared to observations in Figure
\ref{fig:scalez0}. Our observational sample ranges
from local dwarf galaxies to the most actively star-forming galaxies
in the local Universe including objects taken from the THINGS survey
\citep{Leroy2008,Walter2008}, GASS+COLDGASS
\citep{Catinella2010,Saintonge2011}, LITTLE THINGS survey
\citep{Hunter2012,Zhang2012}, \citet{McGaugh2012},
\citet{Boselli2014}, and the ALLSMOG survey \citep{Bothwell2014}. 

We find good agreement between model and observations for the relative and absolute \hi content of galaxies
over the entire mass range probed. The \hi mass of galaxies increases
with stellar mass, but the ratio between
\hi and stellar mass decreases. We find an increasing trend in \h2
mass with stellar mass, in good agreement with observations. The relation
between stellar mass and \h2 mass is steeper than the relation between
stellar mass and cold gas and \hi mass. The inferred \h2-to-stellar mass ratio is flat up to stellar
masses of $M_{\star} \sim 10^{10.5}\,\rm{M}_\odot$ and drops afterwards. Observations suggest a continuous decline in the \h2-to-stellar mass
ratio  with stellar mass, however, the \h2 mass of galaxies
with low stellar masses is
not well constrained ($M_\star < 10^8\,\rm{M}_\odot$). This is partially due to small sample sizes and partially
due to uncertainties in the \h2 mass as estimated through CO for
low-metallicity objects \citep{Schruba2011}.

\begin{figure*}
\includegraphics[width = 0.9\hsize]{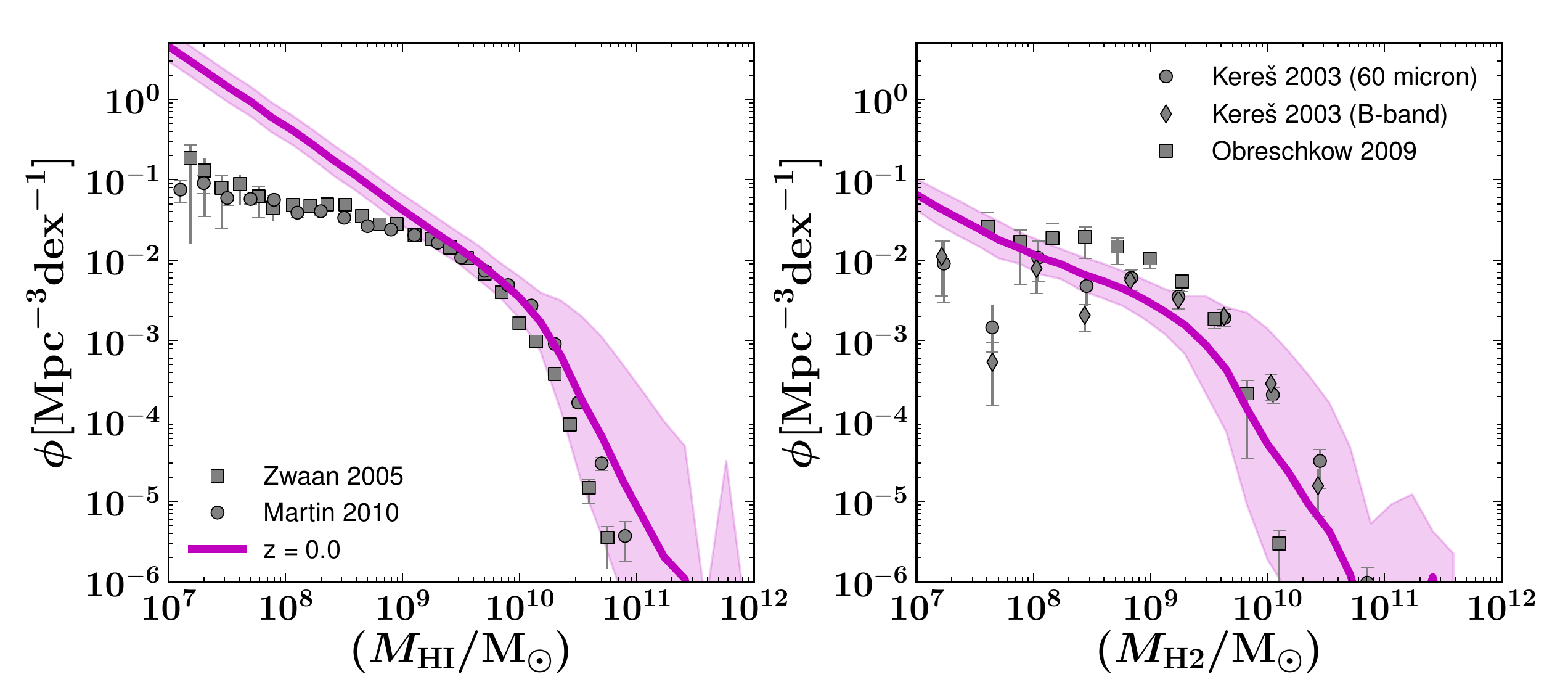}
\caption{\hi (left) and \h2 (right) mass function at $z=0$. Model
  results are compared to literature results from
  \citet{Zwaan2005}, and \citet{Martin2010} for \hi and
  \citet{Keres2003} and \citet{Obreschkow2009} for \h2. The shaded
  regions show the one-sigma posterior distributions.
\label{fig:massfunc_z0}}
\end{figure*}

We compare our inferred local \hi and \h2 mass functions to observations in 
Figure \ref{fig:massfunc_z0}. We reach decent agreement with the
observed \hi mass function for galaxies with \hi masses $M_{\rm
    HI} > 10^9\,\rm{M}_\odot$. Below this mass, our model finds a much
steeper slope in the \hi mass function than suggested by
observations. The galaxies
responsible for this excess have low stellar masses ($M_{\star} <
10^{8.5}\,\rm{M}_\odot$). We will further discuss this in Section \ref{sec:HI_excess}.

We obtain good agreement between the observed and inferred \h2 mass
function over the entire mass range probed. Only
at the knee of the \h2 mass function ($M_{\rm H2} \sim 10^{9.5}$) do we slightly underpredict the number density of
galaxies compared to observations. 

Overall we successfully infer the gas properties
of galaxies with stellar masses larger than $10^{8.5}\,\rm{M}_\odot$. The SHAM adopted in this
work was designed to match the locally observed stellar mass function
and specific star-formation rates of galaxies \citep{Behroozi2013}. Our methodology to estimate the \hi and \h2 masses of local
galaxies was calibrated using a sample partially consisting of local
gas rich objects. Although the combination of these two models does
not necessarily guarantee the degree of agreement found
between our model and observations, the good match is an encouraging
sanity check.

\begin{figure*}
\includegraphics[width = \hsize]{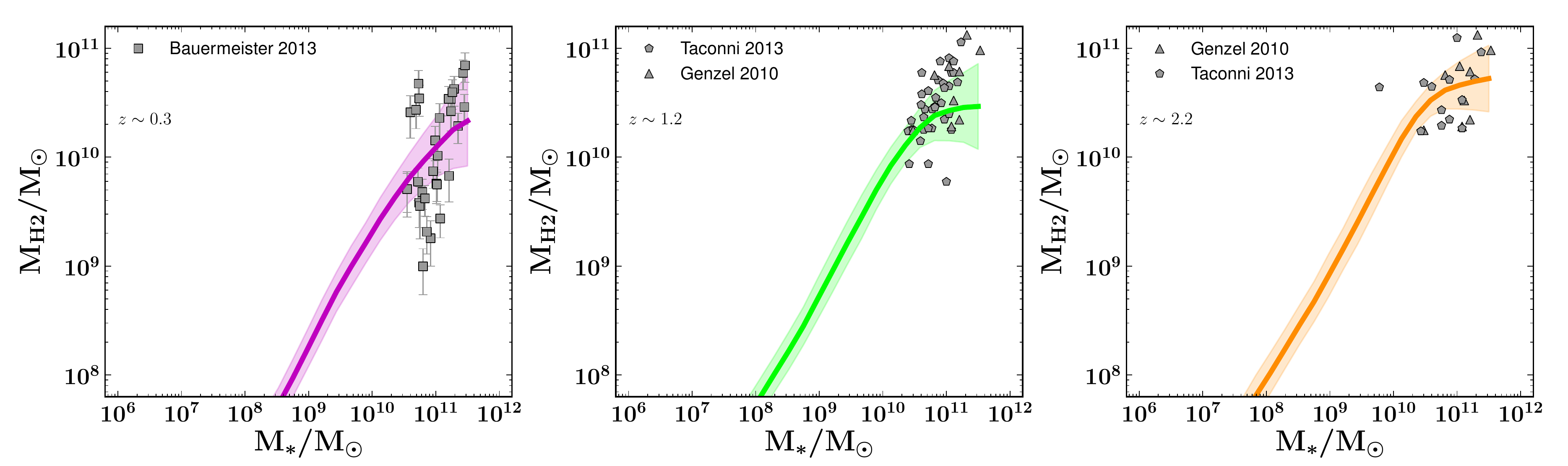}
\caption{\h2 mass as a function of stellar mass at redshifts $z=0.3$,
  $z=1.2$, and $z=2.2$. Shaded regions mark the one-sigma
  distributions of the inferred gas masses. Model results are compared to observations from \citet{Bauermeister2013} ($z=0.3$), \citet{Tacconi2013}
  ($z=1.2$ and $z=2.2$), and the compilation presented in
  \citet{Genzel2010} ($z=1.2$ and $z=2.2$).\label{fig:mstarH2evol}}
\end{figure*}

\begin{figure*}
\includegraphics[width = \hsize]{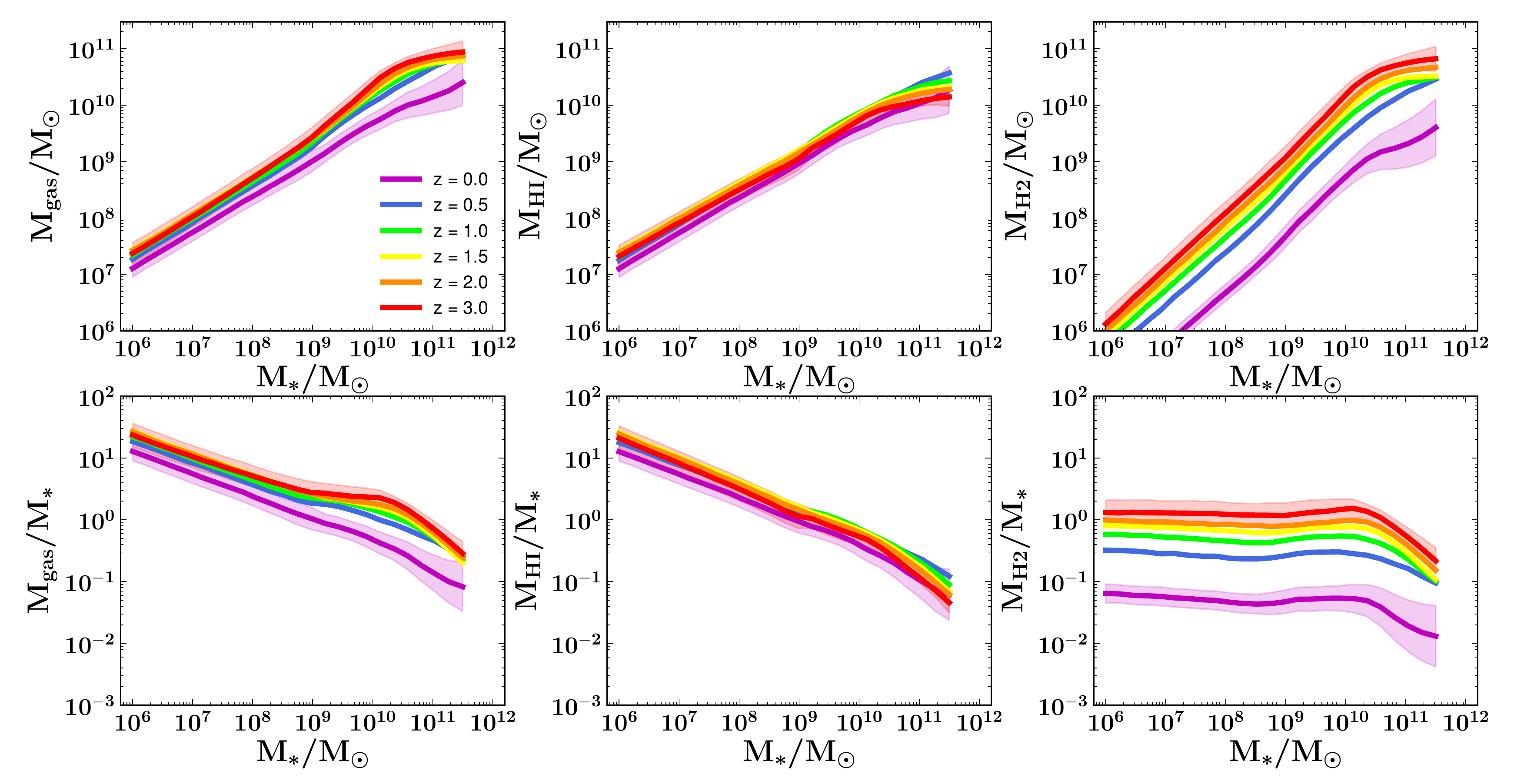}
\caption{Redshift evolution of the cold gas mass (top left), \hi mass (top row center
  column), \h2 mass (top right), gas-to-stellar mass ratio (bottom
  left), \hi-to-stellar mass ratio (bottom row center column), and \h2-to-stellar mass ratio
  (bottom right) of galaxies as a function of stellar mass. Shaded
  regions mark the one-sigma posterior distributions of the inferred gas masses. For clarity we only show the
  one-sigma region at redshifts $z=0$ and $z=3$. The distribution
  at other redshifts is very similar. \label{fig:scaleevol}
}
\end{figure*}

\subsection{Evolution of gas in galaxies}
\label{res:highz}
\subsubsection{Scaling relations}
We present inferred galaxy \h2 masses at redshifts $z\sim 0.3$,
$z\sim 1.2$, and $z\sim 2.2$ in Figure \ref{fig:mstarH2evol}. At these redshifts there are
ample observations from the literature available to compare our model to
\citep{Tacconi2010,Bauermeister2013,Tacconi2013}. We find an
increasing trend in galaxy \h2 mass with 
stellar mass at all redshifts, in good agreement with observations. The relation between
\h2 mass and stellar mass flattens for the most stellar-massive
galaxies ($M_{\star} > 10^{11}\,\rm{M}_\odot$), suggestive of an upper
plateau in \h2 mass. As for the local gas
measures, our methodology was designed to properly reproduce observed
stellar mass functions and SFRs in this redshift range and our
indirect gas measure recipe was calibrated using among others the
Tacconi et al. observations. Nevertheless, it is encouraging that our
model simultaneously reproduces the available gas measures in the local and
high-redshift Universe. 

We show the redshift evolution of the cold gas, \hi and \h2 mass and the
cold gas-, \hi- and \h2-to-stellar mass ratio of galaxies up to $z=3$
in Figure \ref{fig:scaleevol}. The shapes of the presented relations are roughly
constant with time. The cold gas, \hi, and \h2 mass of a galaxy increases and its
gas-to-stellar mass and \hi-to-stellar mass ratios decrease with stellar mass. The
\h2-to-stellar mass ratio remains constant below stellar masses of
$M_{\star} \sim 10^{10 - 10.5}\,\rm{M}_\odot$ and decreases for galaxies with
higher stellar masses, independent of redshift. 

It is striking to
see that as time evolves the normalization of the absolute and
relative \hi and \h2 contents of galaxies show very different
behavior. We find very weak evolution in the \hi content for galaxies
with low stellar masses ($M_{\star} < 10^{9.5}\,\rm{M}_\odot$). Galaxies with
higher stellar masses show a much stronger evolution in their relative
and absolute atomic gas mass. The \hi mass of galaxies increases up to $z=0.5$ and stays constant
at later times. Below $z=1$ the evolution in \hi mass is less than a
factor of three. We find very different evolution in the \h2 mass of
galaxies with time. Galaxies at $z=3$ have roughly ten times as much molecular hydrogen
than local counterparts, independent of stellar mass and the
strongest evolution in \h2 mass is at $z<1$.

The relations between cold gas mass and gas-to-stellar mass
ratios of galaxies and stellar mass are constant with time at $z<0.5$. Only at later
times does the total cold gas content of galaxies rapidly decrease.

\begin{figure*}
\includegraphics[width = 0.9\hsize]{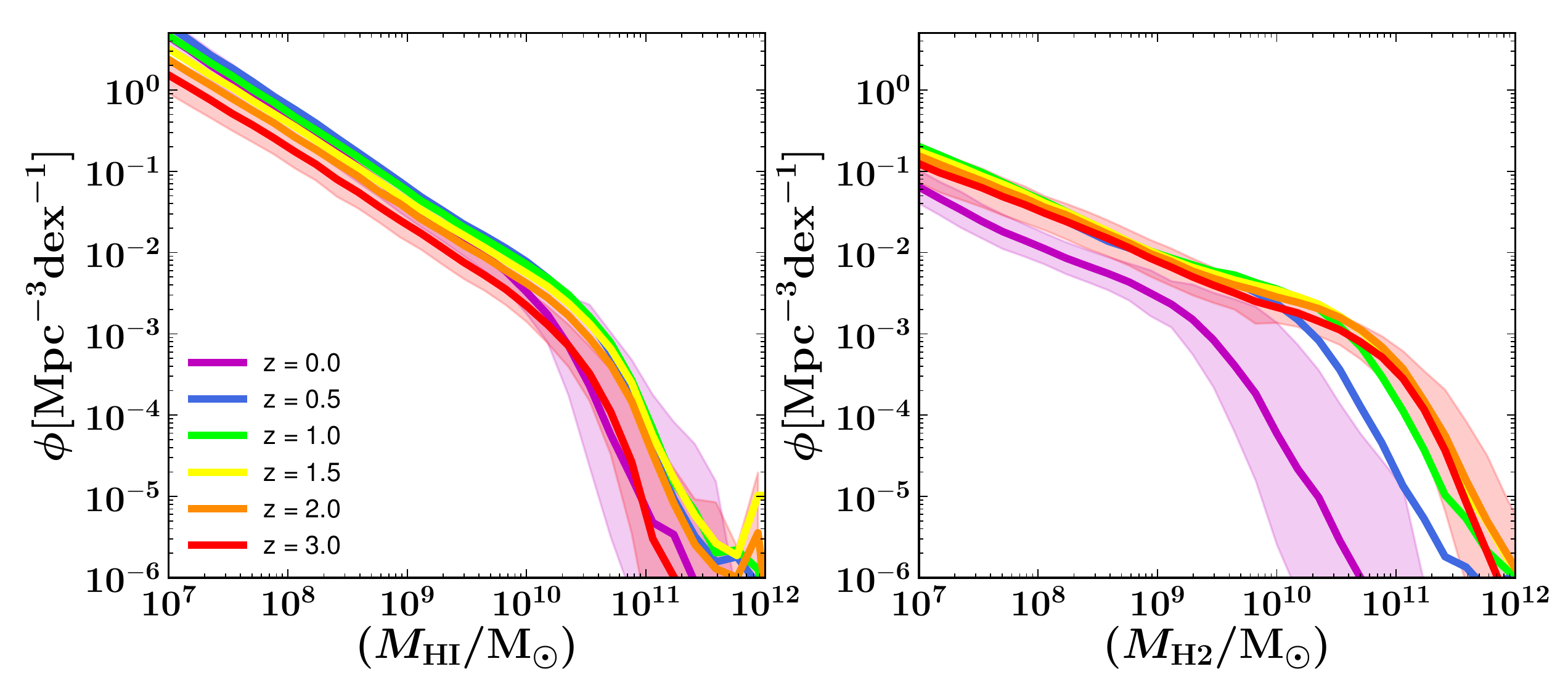}
\caption{Evolution of the \hi (left) and \h2 (right) mass functions
  with redshift. The shaded regions show the one-sigma posterior
  distributions. For clarity we only show the
  one-sigma region at redshifts $z=0$ and $z=3$. The distribution
  at other redshifts is very similar.
\label{fig:massfuncevol}}
\end{figure*}

\begin{figure*}
\includegraphics[width = 0.9\hsize]{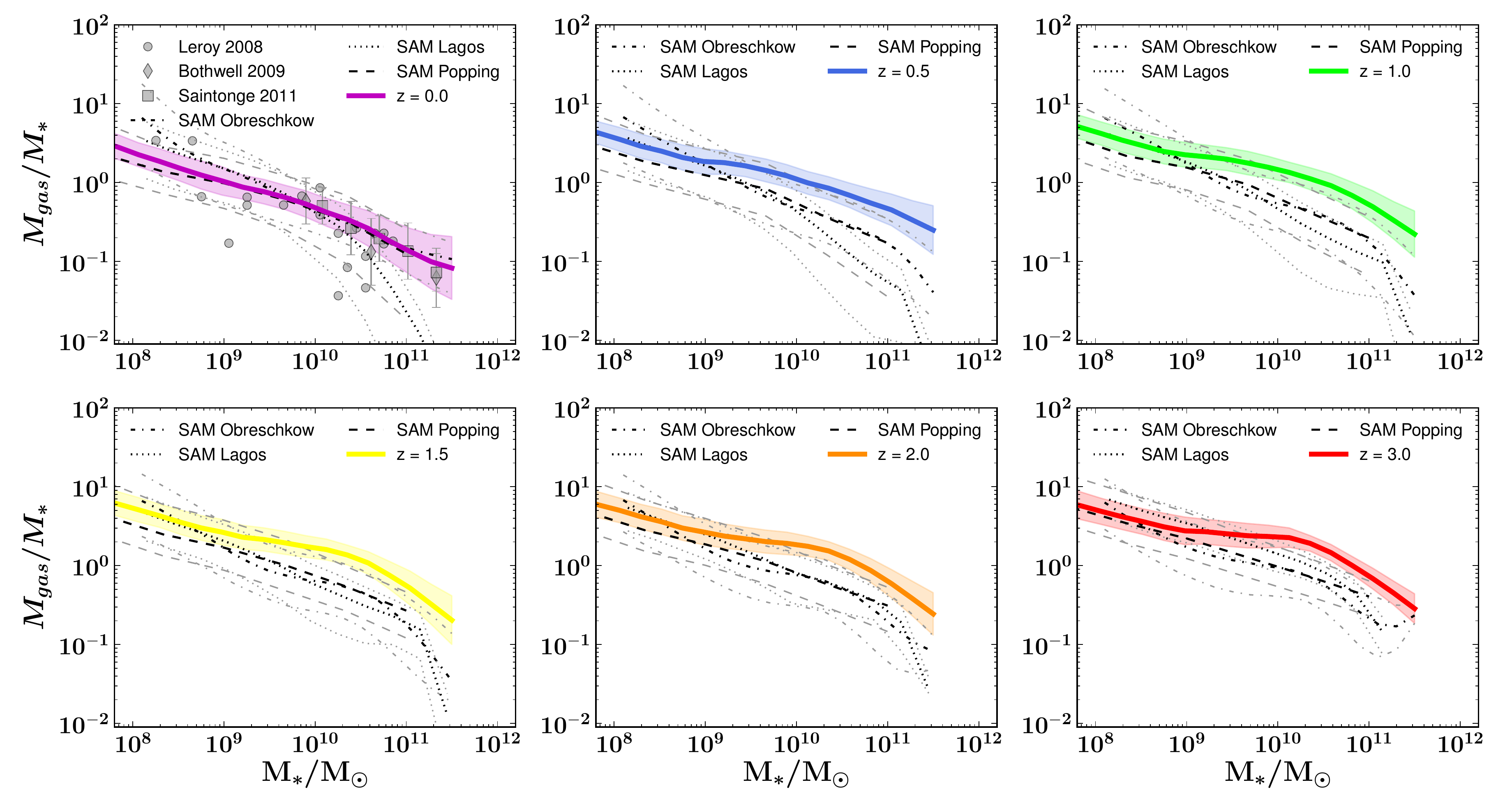}
\caption{Gas fraction of galaxies as a function of stellar mass for
  different redshift bins. Shaded regions mark the one-sigma posterior distributions of the inferred gas fractions. Model results at $z=0$ are compared
  to literature values from
  \citet{Leroy2008}, \citet{Bothwell2009}, and
  \citet{Saintonge2011}. The black dashed, dotted, and dash-dotted
    lines represents the semi-analytic model predictions from
    \citet{Popping2013}, \citet{Lagos2011cosmic_evol}, and \citet{Obreschkow2009_sam}.\label{fig:gasfracevol}}
\end{figure*}
\begin{figure*}
\includegraphics[width = 0.9\hsize]{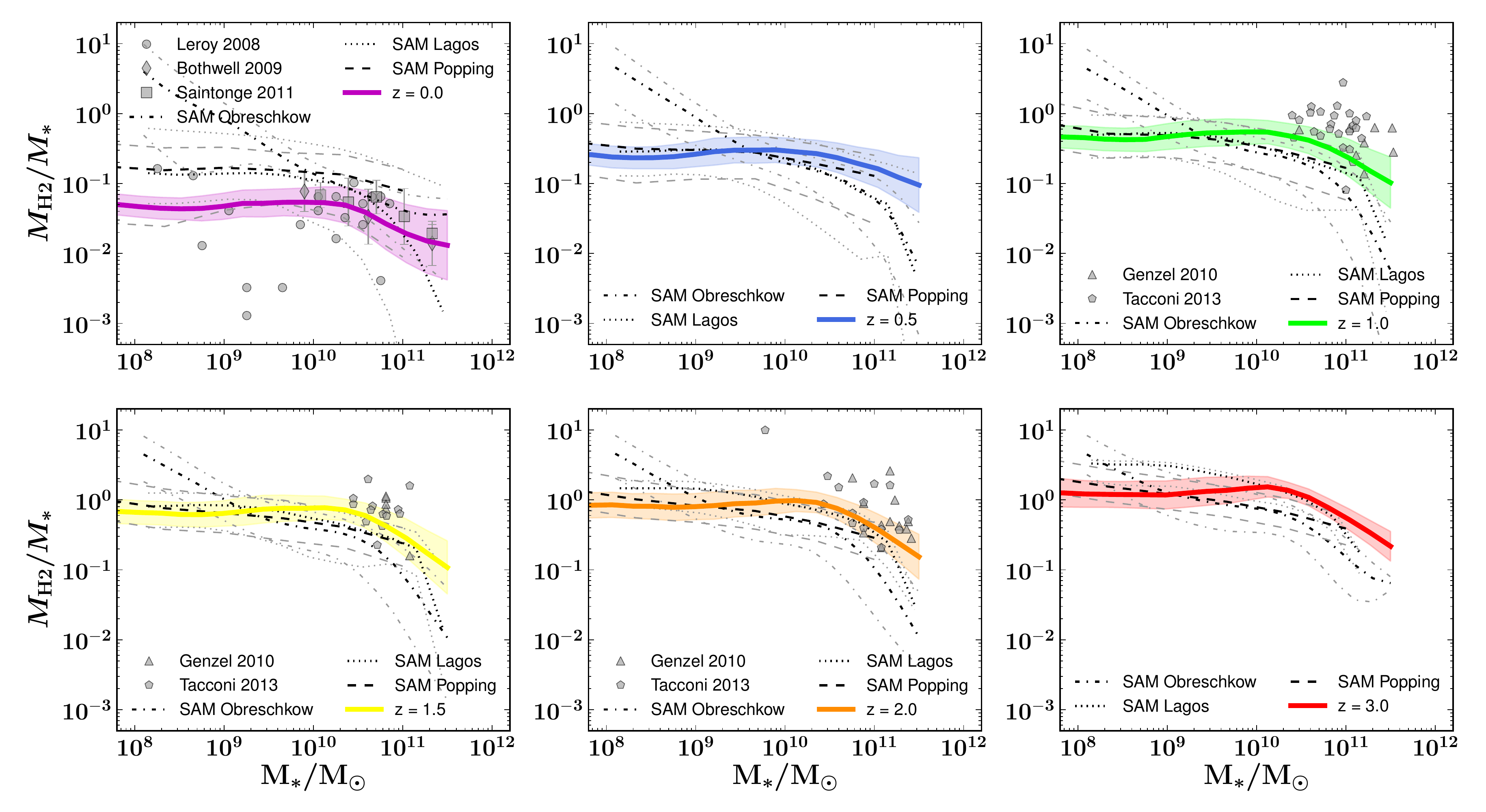}
\caption{Relative \h2 content galaxies (\frach2star) as a function of stellar mass for
  different redshift bins. Shaded regions mark the one-sigma posterior distributions of the inferred gas fractions. Model results are compared
  to literature values from
  \citet{Leroy2008},\citet{Bothwell2009}, \citet{Genzel2010}, \citet{Saintonge2011},
  and \citet{Tacconi2013}. The black dashed, dotted, and dash-dotted
    lines represents the semi-analytic model predictions from
    \citet{Popping2013}, \citet{Lagos2011cosmic_evol}, and \citet{Obreschkow2009_sam}.\label{fig:H2fracevol}}
\end{figure*}

\subsubsection{\hi and \h2 mass function}
We show the redshift evolution of the \hi and \h2 mass functions up to
redshift $z=3$ in Figure \ref{fig:massfuncevol}. At fixed \hi mass the number density of
objects increases up to $z=0.5$ and decreases at later times. 
At low \hi masses ($M_{\rm HI} <
10^{9.5}\,\rm{M}_\odot$) the evolution in \hi mass function with time is less
than a factor three in number density at fixed \hi mass. At higher masses the number density of objects
rapidly increases up to $z=2$, after which the evolution is
weak. Only at $z < 0.5$ does the number density of galaxies with highest
\hi masses drop by an order of magnitude. The evolution in the \hi mass
function is in good agreement with predictions from semi-analytic
models \citep[e.g.,][]{Popping2013}, but in poorer agreement with
prediction using the hydrodynamical simulations presented in \citet{Dave2013} (we discuss this in Section \ref{sec:discussion}). The steep
slope in the low-mass end of the \hi mass function as seen at $z=0$ is present at all
discussed redshifts. We discuss the origin of this steep slope in
Section \ref{sec:HI_excess}. 

Our inferred \h2 mass function has the highest normalization at
redshift $z\sim 2$. Largest \h2 reservoirs ($M_{\rm H2}\sim 10^{11.5-12}\,\rm{M}_\odot$) are also
found at this redshift. These two results
(highest number and most massive objects) coincide with the peak of
the SFR density distribution of the Universe. The evolution in the \h2
mass function at redshifts $1 < z < 3$ is weak. We find a
much stronger evolution in the \h2 mass function at lower
redshifts. The rapid decline in number density of galaxies with
$M_{\rm H2} > 10^{10}\,\rm{M}_\odot$ starts at $z=1$, whereas the
decline in number density of less massive galaxies starts at
$z<0.5$. This suggests that galaxies with high \h2 masses lose their molecular gas at earlier times than less molecule rich galaxies.

It is intriguing to see that the evolution in the \hi and \h2 mass
functions do not go hand-in-hand. The strongest evolution in \h2 is
observed at $z<1$, where the cosmic SFR density is plummeting,
whereas the \hi mass function is much more
constant in this redshift regime. Although the number of massive total
(\hi + \h2) cold gas reservoirs changes (especially at $z<1$), this
does not automatically mean that \hi and \h2 independently follow the
same trend. 

\begin{figure*}
\includegraphics[width = 1\hsize]{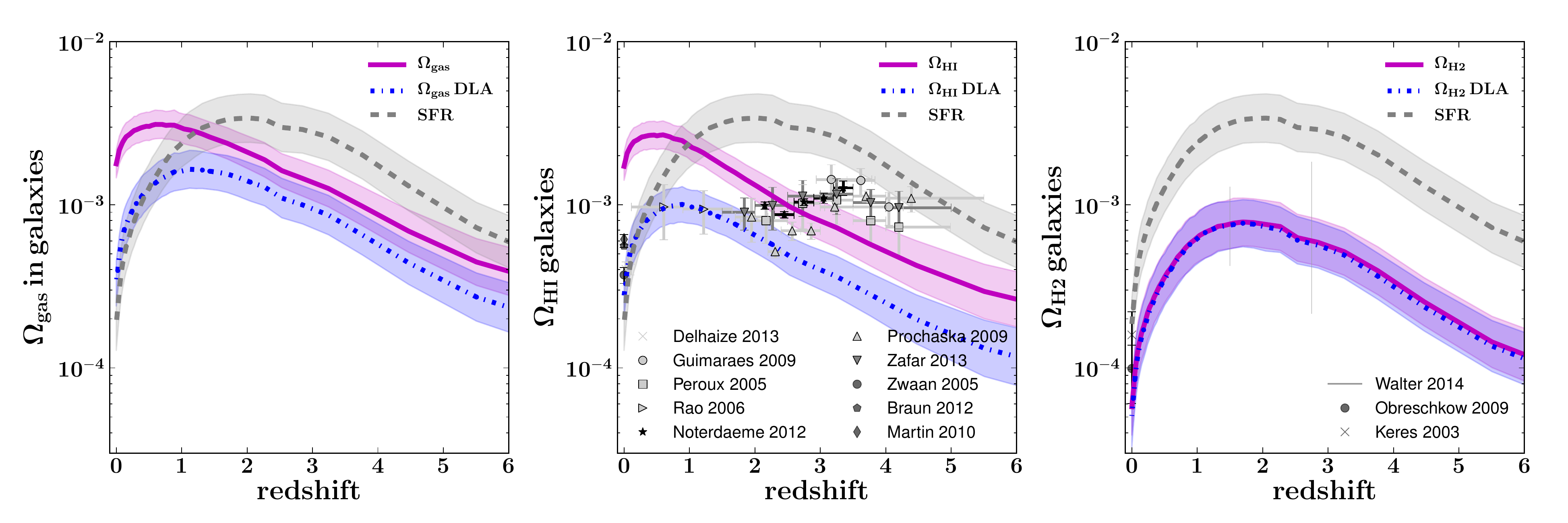}
\caption{The cosmic comoving density, in units of the critical
  density, of cold gas (\hi + \h2; top panel), \hi (middle) and \h2 (bottom) as a
  function of redshift. Observations of \citet{Peroux2005},
  \citet{Rao2006}, \citet{Guimaraes2009}, \citet{Prochaska2009}, \citet{Delhaize2013} are
  overplotted in light gray. Dark gray observations are by
  \citet{Zafar2013} and observations of local galaxies \citep{Keres2003,Zwaan2005,Martin2010,Obreschkow2009,Braun2012} are
  overplotted in black. Grey bars mark the observational constraint on
  $\Omega_{\rm H2}$ from \citet{Walter2014}. The dotted grey line marks the evolution of
  the SFR density of the Universe in $\rm{M}_\odot\,{\rm
    yr}^{-1}\,\rm{Mpc}^{-3}$, scaled down by a factor of 50. The shaded
  regions show the one-sigma posterior distributions. \label{fig:Omega_evolution}}
\end{figure*}

\subsubsection{Galaxy gas fractions}
We present the gas fraction and relative \h2 content of galaxies as a
function of stellar mass in Figures \ref{fig:gasfracevol} and
\ref{fig:H2fracevol}. Gas fractions are plotted at redshifts running
from $z=0$ to $z=3$ and compared with direct observations and predictions based on the
semi-analytic models by
\citet{Obreschkow2009_sam}, \citet{Lagos2011cosmic_evol}, and \citet{Popping2013}. The inferred gas masses
can act as constraints for theoretical models and prescriptions that describe the baryonic cycle of matter within galaxies.

The cold gas fraction of a galaxy decreases with increasing stellar
mass. At $z>1$ we find a shallow slope in gas fraction for galaxies with
$M_{\star} < 10^{10}\,\rm{M}_\odot$ followed by a rapid drop for galaxies with
higher stellar masses. At lower redshifts
galaxy gas fractions monotonically increase with stellar
mass. The evolution in the normalization of galaxy gas fraction with time is weak. Most prominent evolution is found at stellar masses around $10^{10}\,\rm{M}_\odot$ and below $z = 0.5$.

The evolution in \frach2star is much stronger than found for
$f_{\rm gas}$ (Figure \ref{fig:H2fracevol}). At $z=3$, \frach2star reaches values of $\sim 2$,
whereas it less than 0.1 at $z=0$. Although at $z=0$
low-stellar mass galaxies are dominated by their cold gas content, the \h2
fraction of this gas is very low. At $z=3$, on the other hand, most
of the cold gas is molecular. We find a peak in \frach2star for
galaxies with stellar masses near $10^{10}\,\rm{M}_\odot$ at all
redshifts. There is a minor increase with stellar
mass below this characteristic mass, whereas we find a rapid drop in
\frach2star above this characteristic mass. This suggests galaxies are
most efficient in forming \h2 and transforming cold gas into stars
near $M_{\star} = 10^{10}\,\rm{M}_\odot$. We will explore which haloes host
these galaxies and what light this sheds on the stellar build-up of
galaxies in Section \ref{sec:gashalo}.

We compare our findings for \frach2star to literature values
presented in \citet{Tacconi2013} and the compilation from
\citet{Genzel2010} (obtained through CO). Direct observations of \frach2star are within the
scatter of our inferred \h2 masses, although at the high side compared to our
mean trends. We ascribe this to the strongly biased nature of the
observations at these redshifts towards actively star-forming (and therefore \h2 rich)
galaxies. Nevertheless the rough agreement between observations and
our results for these stellar massive galaxies is very encouraging.

The presented trends provide  constraints for theoretical
models including a detailed prescription for galaxy cold gas
content. These constraints have the potential to break degeneracies
between different physical recipes included in models. Unlike indirect gas estimates from large optical surveys
\citep[Popping et al. in prep.]{Popping2012}, the work presented here is not
biased by selection criteria and represents a complete sample.  Semi-analytic model predicts a similar shape in the trend
between gas fractions and \frach2star with stellar mass and with
redshift as our semi-empirical model. The normalization
of the trends at $M_{\star} > 10^{10}\,\rm{M}_\odot$ predicted by semi-analytic models at $z \geq 0.5$ is systematically below the results presented in
this work. The discrepancy is especially obvious in the gas fractions in the
redshift regime $1 < z < 2.5$. This suggests that in semi-analytic
models galaxies have expelled and/or consumed more of their cold
gas. A similar dearth of molecular hydrogen in this redshift regime was found in a comparison between
semi-analytic model predictions and indirect \h2 estimates taken from
the COSMOS sample \citep{Lu2013,Popping2013,White2014}. This is
likely coupled to the low-mass galaxy problem, where galaxies in
theoretical models form their stars and consume their gas too early.

\begin{figure}
\includegraphics[width = 1\hsize]{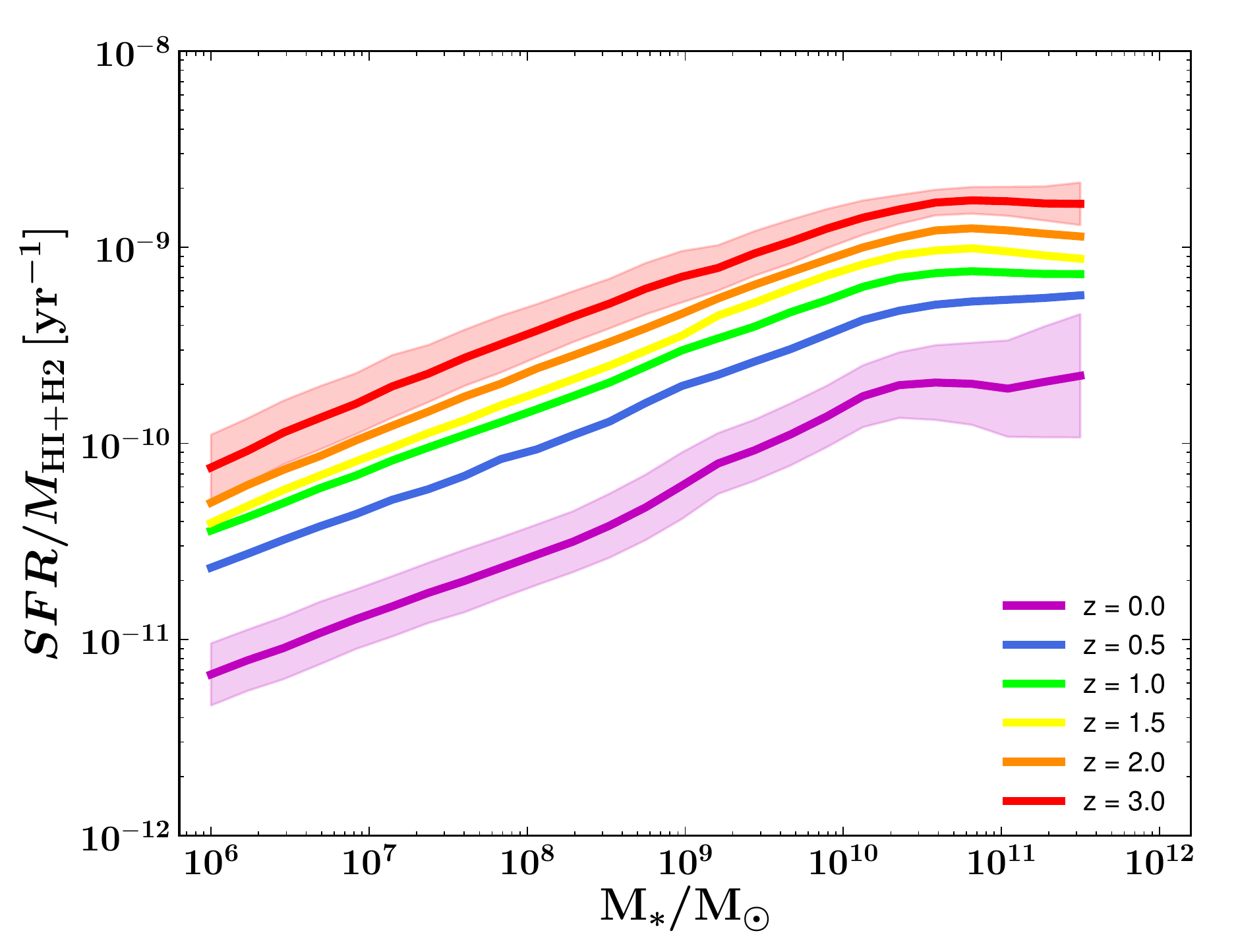}
\caption{Redshift evolution of the star-formation efficiency ($\rm{SFE} \equiv
  \rm{SFR}/M_{\rm HI + H2}$) of galaxies as a function of stellar
  mass. Solid lines mark the mean trends and shaded colors mark the one-sigma
  distribution of the inferred gas masses. For clarity we only show the
  one-sigma regions at redshifts $z=0$ and $z=3$. The distribution
  at other redshifts is very similar.
\label{fig:SFE}}
\end{figure}

\begin{figure}
\includegraphics[width = 1\hsize]{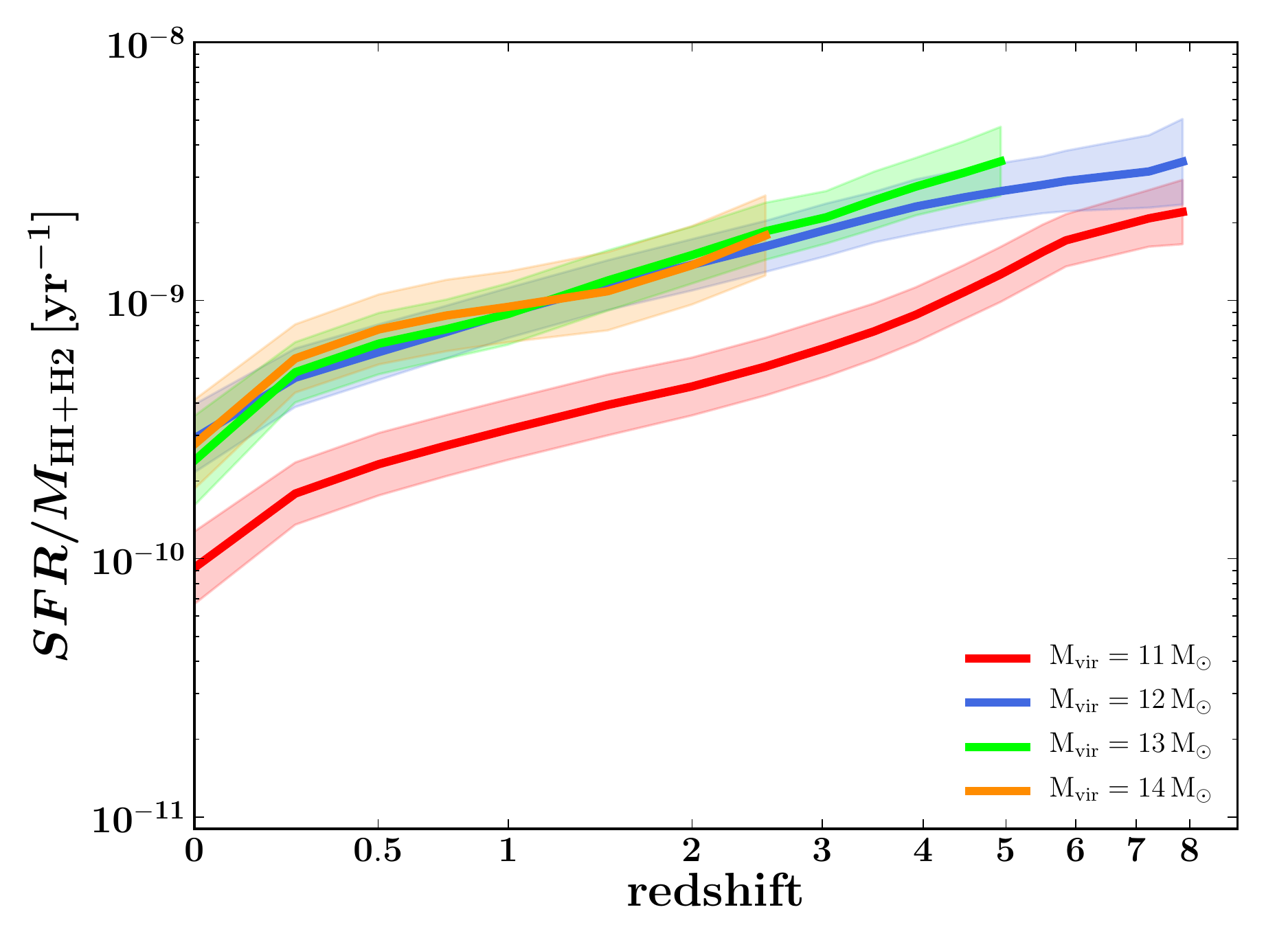}
\caption{ The star-formation efficiency ($\rm{SFE} \equiv
  \rm{SFR}/M_{\rm HI + H2}$) of galaxies as a function of redshift in different halo mass bins. Shaded regions mark the one-sigma
  distributions of the star-formation efficiencies. \label{fig:SFE_z}}
\end{figure}

\subsubsection{Cosmic gas density in galaxies}
We present our inferences for the total cold gas (\hi + \h2), \hi, and \h2 cosmic density of the
Universe as a function of time in Figure \ref{fig:Omega_evolution}. We compare our results to local \hi
and \h2 densities \citep{Keres2003,Zwaan2005,Obreschkow2009,Martin2010,Braun2012} and
higher redshift estimates of the \h2 \citep{Walter2014} and \hi density obtained through stacking
\citep{Delhaize2013} and from DLAs \citep[e.g.,][]{Peroux2005,
  Rao2006,Guimaraes2009,Prochaska2009,Zafar2013}. 

We find a gradual increase in the total cold gas cosmic density with
redshifts up to $z \sim 1$ followed by a rapid drop at lower
redshifts. A similar trend is found for the \hi cosmic density, which
peaks at $z \sim 0.5$. The evolution in $\Omega_{\rm HI}$ is relatively weak in
the redshift regime $0 < z < 2$. The \h2 cosmic density on the other hand increases gradually
up to $z\sim 1.5$ and shows a rapid drop at lower redshifts, closely
following the evolution of the cosmic SFR. The inferred \h2 cosmic density of
gas in galaxies never exceeds the \hi cosmic density.

We reach good agreement
with the constraints on $\Omega_{\rm H2}$ at redshifts $z=1.5$ and
$z=2.75$ \citep{Walter2014}, supporting an increase in the cosmic \h2
density. The observational constraints are unfortunately not tight enough to separate
between a peak, a flat, or declining evolution in  $\Omega_{\rm H2}$
at $z > 2$. Our approach overpredicts the observational estimates of $\Omega_{\rm
  HI}$ at $z < 2$ and underpredicts $\Omega_{\rm
  HI}$ at redshifts $z > 3$. In the low redshift regime this is due
to the high number of galaxies with low \hi masses, in poor agreement
with the observed \hi mass function at $z=0$. We warn the reader that in this work we
only probe the cold gas that is associated to the galaxy disc (down to
a hydrogen column density of $N_{H}
= 10^{18}\,\rm{cm}^{-2}$), ignoring DLAs that may arise from intergalactic gas  in cold streams
or filaments at $z>1.5-2$ \citep{Berry2013}. As a comparison we include inferred cosmic densities when only taking gas above the
DLA column density into account (i.e. $N_{\rm H} > 2\times
10^{20}\,\rm{cm}^{-2}$). The resulting cosmic densities for cold gas
and \hi decrease by a factor of $\sim4$, whereas this selection does not affect
the \h2 cosmic density. Our predictions for the \hi cosmic densities
are now in good agreement with observations out to $z\approx
2$, but at higher redshifts our inferred cosmic \hi densities are lower
than the observed densities.

\begin{figure*}
\includegraphics[width = 1\hsize]{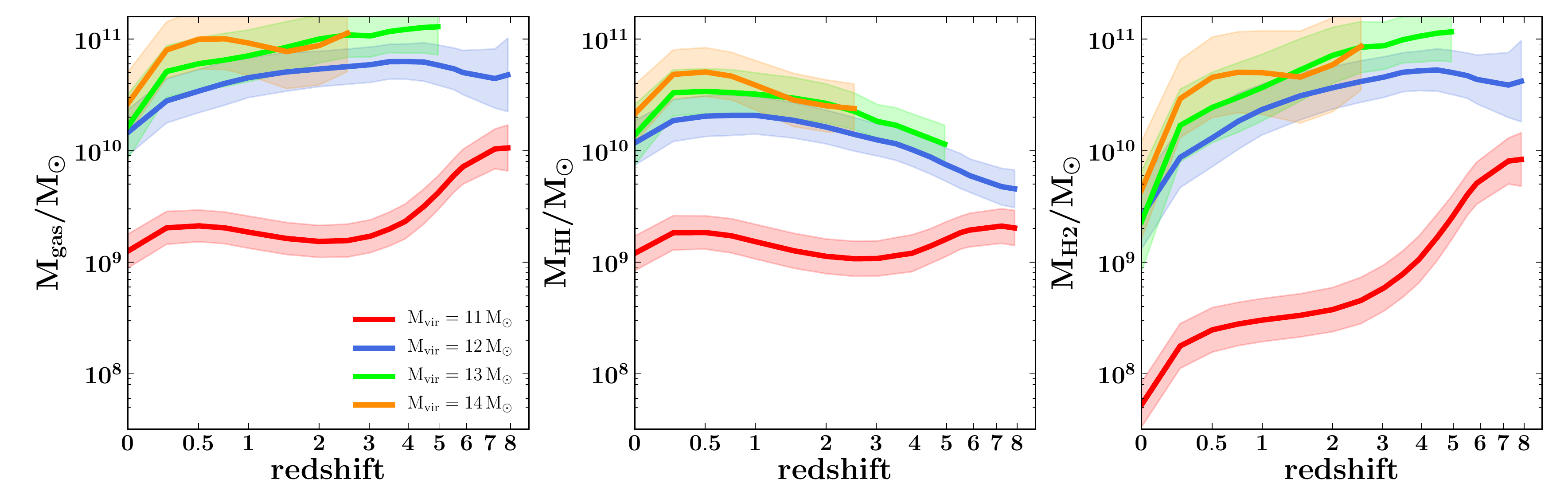}
\caption{Average cold gas mass (\hi + \h2; left
  panel), \hi mass (middle panel), and \h2 mass (right panel) as a
  function of redshift for
  galaxies in haloes at fixed halo mass. Shaded regions mark the
  one-sigma posterior distributions of the inferred gas masses.\label{fig:mgasmhalo}}
\end{figure*}

\subsubsection{Star-formation efficiency}
The star-formation efficiency (here defined as $\rm{SFE}\equiv
\rm{SFR}/M_{\rm HI + H2}$) measures the ability of a galaxy to
transform cold gas into stars.\footnote{This is different from a commonly
  used definition for star-formation efficiency; the fragment of gas mass
   converted into stars} We present the SFE of galaxies as a
function of stellar mass and redshift in Figure \ref{fig:SFE}. The SFE
of galaxies increases with stellar mass, independent of
redshift. Although the relative gas content of galaxies decreases
with stellar mass (Figure \ref{fig:gasfracevol}), the ability of the
gas to form stars increases. This is largely set by a decrease in the
\hi content of galaxies with increasing stellar mass. At fixed stellar mass, the SFE of
galaxies decreases with time, resembling the evolution of the \h2
content of galaxies (Figure \ref{fig:scaleevol}; note, however, that
in our model the \h2 mass of a galaxy is largely controlled by its SFR).

There is a plateau in the SFE of galaxies at stellar masses
$M_{\star} > 10^{10}\,\rm{M}_\odot$, independent of redshift. At fixed
redshift, star-forming galaxies have reached their maximum SFE at this
characteristic stellar mass. It is at the same characteristic stellar
mass that the relative \h2 content of galaxies, \frach2star, rapidly
drops. 
We find a similar
constancy in the SFE of galaxies as a function of their redshift in
different bins of halo mass (Figure
\ref{fig:SFE_z}). At fixed redshift the SFE of star-forming galaxies
is constant as a function of halo mass for galaxies residing in haloes
more massive than $10^{12}\,\rm{M}_\odot$. The SFE of galaxies
decreases with time for a fixed host halo mass. The constancy of SFE
at fixed redshift is in agreement with the plateau in SFE we found for
galaxies with stellar masses larger than $10^{10}\,\rm{M}_\odot$.

\begin{figure*}
\includegraphics[width = \hsize]{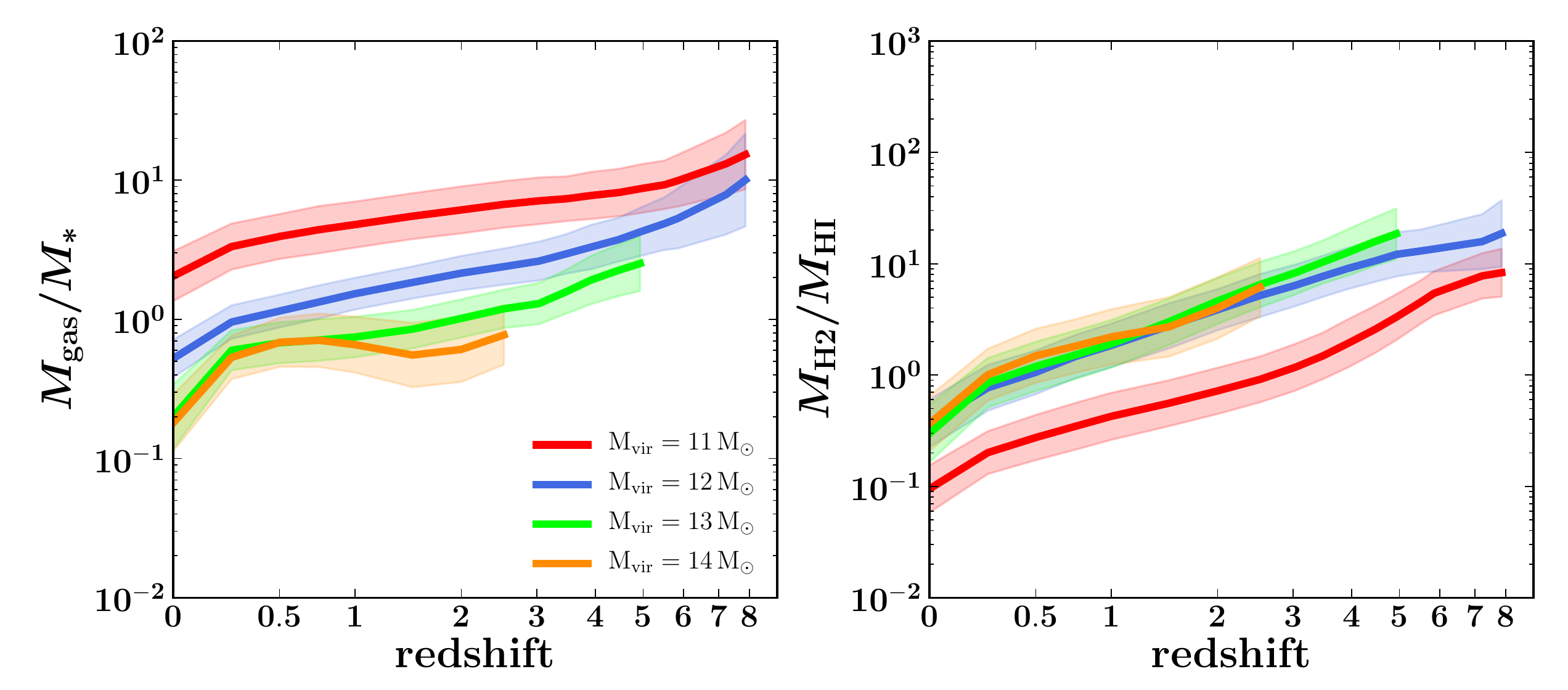}
\caption{Cold gas-to-stellar mass ratio (left) and \h2-to-\hi mass
  ratio (right) of galaxies as a function of redshift in different
  halo mass bins. Shaded regions mark the one-sigma posterior
  distributions of the inferred gas masses. \label{fig:fgasmhalo}}
\end{figure*}
\begin{figure*}
\includegraphics[width = 0.9\hsize]{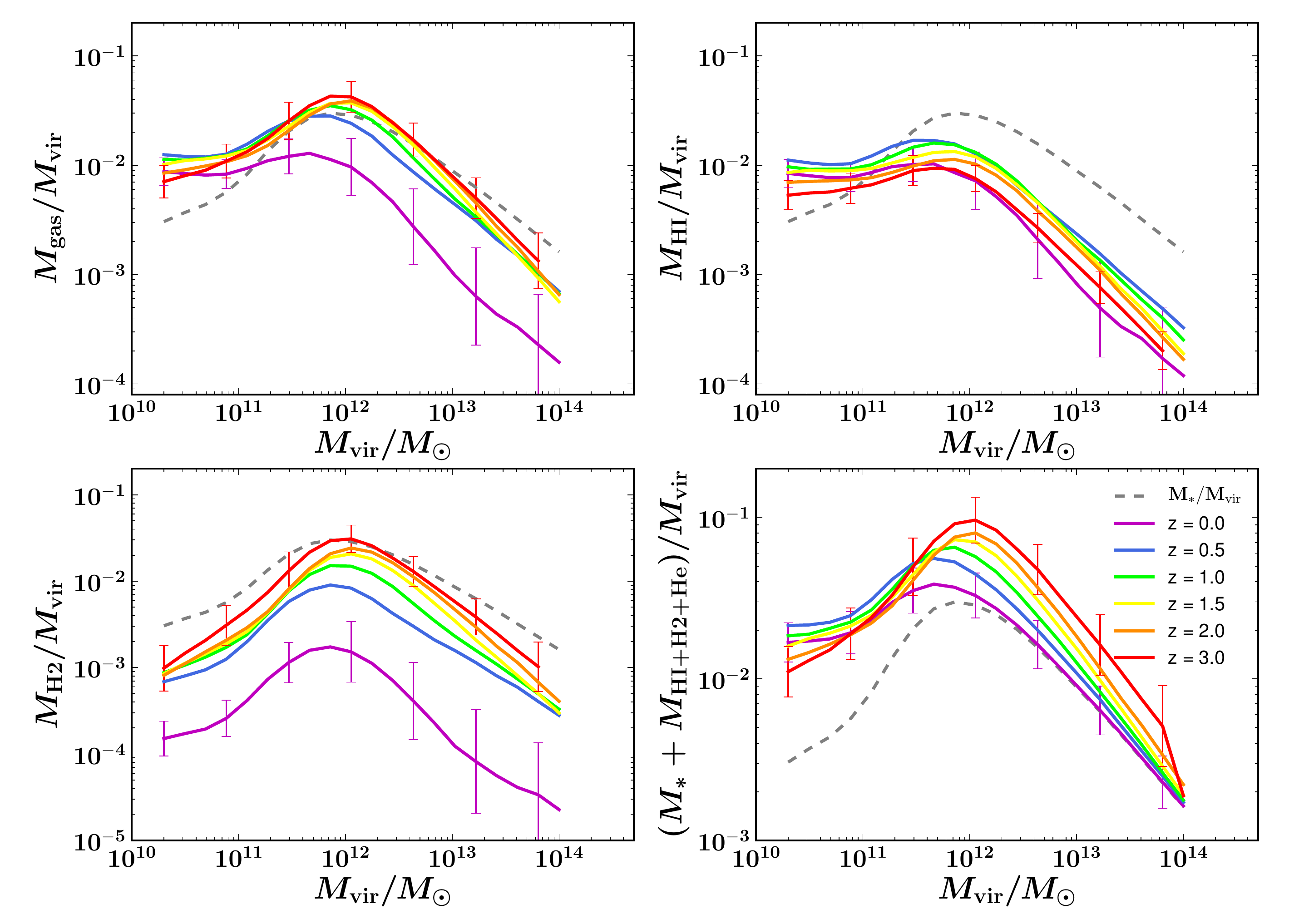}
\caption{Relative cold gas content (\hi + \h2; upper left panel), \hi
  content (upper right panel), \h2 content (lower left panel), and combined gas
  and stellar content including Helium (lower right panel) as a function of halo mass
  for different redshift bins. Errorbars mark the one-sigma posterior distributions of the inferred gas masses. The dotted grey line marks the stellar-to-halo
  mass ratio ($M_{\star}/M_{\rm vir}$) of galaxies at $z=0$.\label{fig:gashaloratio}}
\end{figure*}

\begin{figure*}
\includegraphics[width = 0.9\hsize]{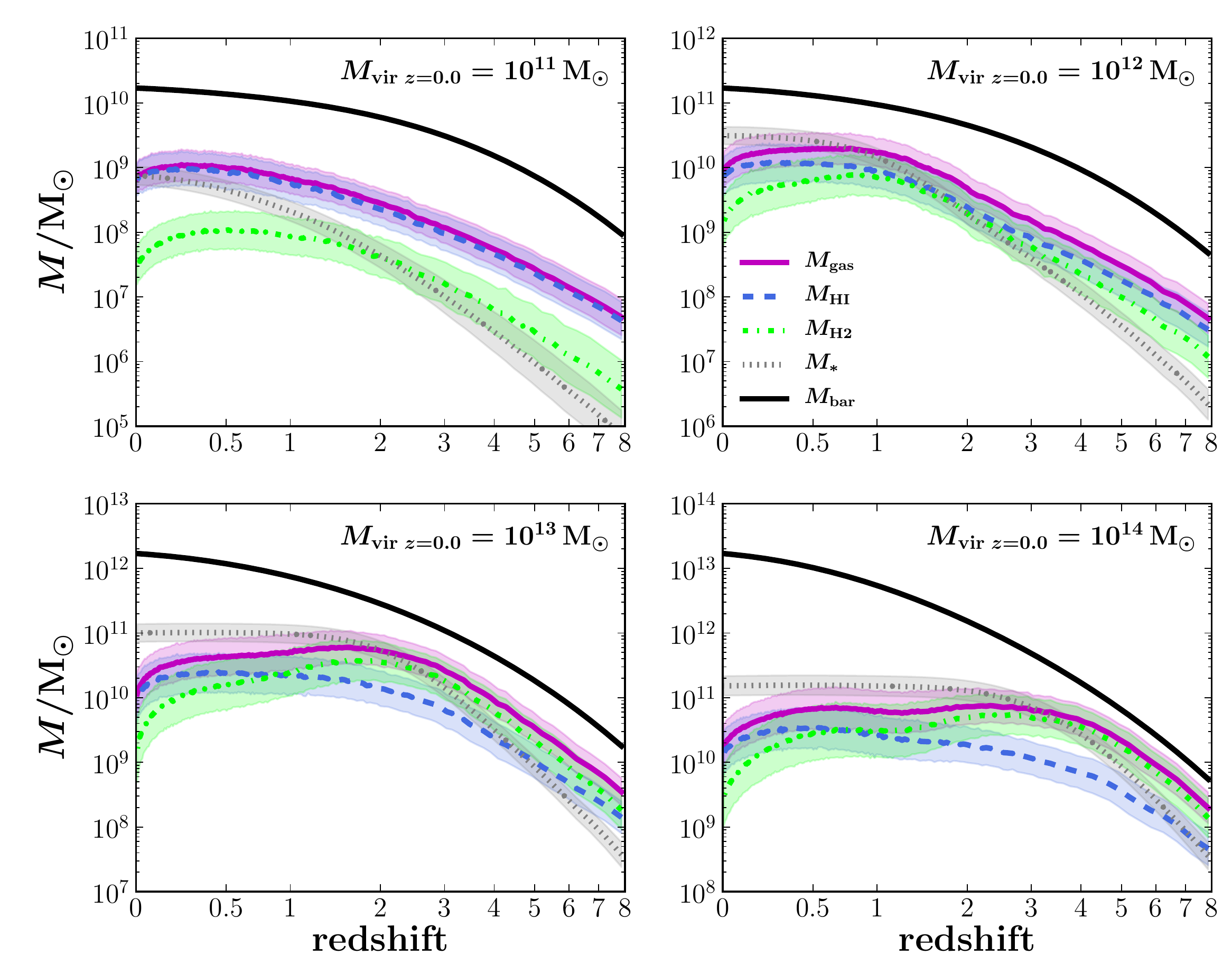}
\caption{Time evolution of the total cold gas (solid purple), \hi
  (dashed blue), \h2 (dashed dotted green), stellar
  (dashed grey) and mass of baryons in the halo (solid black) for haloes as a
  function of their $z=0$ halo mass for the median mass accretion history. The shaded regions show the one-sigma posterior
  distributions.\label{fig:halo_evol}}
\end{figure*}

\subsection{Cold gas content of dark matter haloes}
\label{sec:gashalo}
In this section we present the gas content of galaxies as a function of
their host halo mass. We warn the reader that we only infer gas masses
for star-forming galaxies and do not account for cold gas in
early-type galaxies.

We show derived cold gas, \hi, and \h2 masses as a function of halo
mass and redshift in Figure \ref{fig:mgasmhalo}. We find that the
evolution of the total cold gas content of galaxies is very different as a function of their host halo mass. The gas content of galaxies residing in haloes with $M_{\rm
    h} = 10^{14}\,\rm{M}_\odot$ increases up to $z = 0.5$ after
which it rapidly decreases. Less massive haloes show a weak decrease in their cold gas content with
time. The gas content of galaxies residing in haloes with mass $M_{\rm
    h} = 10^{11}\,M_{\odot}$ rapidly decreases with time until $z\sim 3$,
after which it stays relatively constant.

Much stronger evolution is found for the \hi and \h2 mass
of galaxies. Independent of halo mass the \hi mass increases with
time up to $z\sim 0.5$ and decreases at lower redshifts. The most rapid
increase in atomic gas mass is found for galaxies residing in haloes with masses larger
than $10^{12}\,\rm{M}_\odot$. We find a decrease in the \h2 mass of
galaxies at all redshifts. For galaxies residing in haloes with masses
$M_{\rm vir} \geq 10^{12}\,\rm{M}_\odot$ most rapid decrease is found at redshifts
$z<1$. For galaxies residing in less massive
haloes the decrease in \h2 mass is most rapid at $z>2$ and $z<0.5$.

We find minor differences in the absolute amount of cold gas, \hi
and \h2  in haloes more massive than $M_{\rm
    h}= 10^{12}\,\rm{M}_\odot$. Although in the most massive haloes many more
baryons are potentially available to cool
down onto their central galaxy, these objects
have only twice as large cold gas reservoirs than other
galaxies. The physical processes acting on the cold gas prevent
galaxies to build up gas reservoirs much larger
than on average a few times $10^{10}$ solar masses.

In Figure \ref{fig:fgasmhalo} we show the relative gas and \h2 content
of galaxies as a function of time at fixed halo
mass. Galaxies residing in low-mass haloes  ($10^{11}\,\rm{M}_\odot$) are
dominated by their cold gas content. There is only weak evolution in
the gas-to-stellar mass ratio for these galaxies. We find a gradual
decrease in relative cold gas content with time for more massive haloes. We
find a minor increase in cold gas content until $z=0.5$ followed by a
decline for the most massive haloes probed. Independent of host halo
mass the relative \h2 fraction of the cold gas in galaxies decreases rapidly with time. At $z > 3$ nearly all the cold gas in
haloes is molecular, whereas at $z=0$ it is less than 25 percent. The
relative \h2 fraction is almost constant for haloes with masses of $M_{\rm
    h} \geq 10^{12}\,\rm{M}_\odot$. Despite the cold gas reservoir being
relatively smaller for the highest mass haloes, this does not affect
the ability of the cold gas to form molecules.

We show the ratio of cold gas, \hi, \h2, and total baryonic content in
the disc ($M_{\star} + M_{\rm HI} + M_{\rm H2}$; including a 1.36
contribution for Helium) with halo mass as a function of halo mass
at different redshifts in Figure \ref{fig:gashaloratio}. These ratios
provide insight into the fraction of baryons stored in
cold gas and/or stars in the disc of central galaxies and the relative
amount of baryons in haloes that may participate
in the formation of stars. 

There is a peak in the gas-to-halo mass (\hi and \h2 combined) and in the
\h2-to-halo mass ratio around $M_{\rm vir} =
10^{12}\,\rm{M}_\odot$ at all redshifts. The exact location of this characteristic halo mass moves towards higher masses with increasing
redshift, resembling the characteristic halo mass at which the
stellar-to-halo mass ratio peaks \citep{Behroozi2013}. Below $M_{\rm
  vir} = 10^{12}\,\rm{M}_\odot$ the slope of the \h2-to-halo mass
ratio is much steeper than it is for the \hi-to-halo mass ratio (and
cold gas mass). The drop in gas-to-halo mass ratio for haloes more massive than $10^{12}\,\rm{M}_\odot$ is similar for atomic and molecular
hydrogen. The strong peak in the \h2-to-halo mass ratio suggests that
there is a favorable range of halo masses at which galaxies
host relatively largest molecular gas reservoirs out of which new
stars may form.
 
The normalization of the different relations presented in
Figure \ref{fig:gashaloratio} changes strongly with redshift. The
relative cold gas content of a halo decreases by approximately a
factor of three from $z=3$ to $z=0$. The relative \hi content of a halo grows up to
$z=0.5$ and mildly decreases at lower redshift. The relative \h2 content of a halo is highest at $z=3$ and decreases by more
than a full order of magnitude afterwards. 

We present the total baryonic
content (stars combined with \hi and \h2) in the central galaxy of a halo in the bottom-right panel of Figure
\ref{fig:gashaloratio}. At $z=0$ the ratio between the total
baryonic mass in the galaxy and the halo mass peaks at $M_{\rm vir} \approx 3 \times
10^{11}\,\rm{M}_\odot$. This peak moves towards higher masses
with increasing redshift. The absolute value in the peak of the
fraction of baryons of a halo that are stored in the galaxy decreases from $z=3$ to $z=0$ by less than half a dex. Only
in low mass haloes with masses of a few times $10^{11}$ solar masses
and less does the baryonic content of a halo stored in the central galaxy peak at $z=0.5$, driven by large \hi reservoirs. There is
no evolution in the baryonic mass ratio for galaxies residing in haloes
more massive than $10^{13}\,\rm{M}_\odot$. Galaxies in these haloes have low gas
fractions and are dominated by their stellar content. 

Haloes with masses of a few times $10^{11}$ solar masses are most
efficient stroring their baryons in stars and cold gas within the
central galaxy. At
$z=3$
roughly 30 percent of the available baryons are locked up in stars and
cold gas in the central galaxy. At $z=0$ this number decreased to
roughly 17 percent. The ability of haloes to store their baryons in
stars and cold gas is poorer at lower and higher halo masses. These
results suggest that the low baryon conversion efficiency of haloes
is driven by the inability of haloes to build large gas reservoirs in
galaxy discs.

We present the time evolution of the total cold gas content and \hi
and \h2 mass of galaxies as a function of their $z=0$ halo mass in Figure
\ref{fig:halo_evol}. For massive haloes ($M_{\rm vir} \geq
10^{13}\,\rm{M}_\odot$) the cold gas content increases until the haloes have reached a baryonic mass of approximately $3$--$4\times
10^{11}\,\rm{M}_\odot$ (which translates to a virial mass of roughly
$10^{12}\,\rm{M}_\odot$). At lower redshifts the cold gas content of
galaxies decreases gradually. Although the gas
content of central galaxies decreases, the baryonic content of the halo
keeps growing. For less massive
haloes the cold gas content increases up to $z=1$ ($M_{\rm
  vir}\,_{z=0} = 10^{12}\,\rm{M}_\odot$) or even until almost present
day ($M_{\rm vir}\,_{z=0} = 10^{11}\,\rm{M}_\odot$). 

A close
inspection reveals that the baryonic mass of the host halo at the time that
the galaxies have gained largest gas reservoir increases with more
than an order of magnitude with increasing $z=0$ halo mass ($1.5\times
10^{10}\,\rm{M}_\odot$ for $M_{\rm vir}\,_{z=0} =
10^{11}\,\rm{M}_\odot$ and $5.7\times
10^{11}\,\rm{M}_\odot$ for $M_{\rm vir}\,_{z=0} =
10^{14}\,\rm{M}_\odot$). Furthermore, the gas content increases more
rapidly in galaxies that end up in massive haloes ($M_{\rm vir}\,_{z=0} >
10^{13}\,\rm{M}_\odot$) than in galaxies that end up in less massive
haloes. The increase in cold gas mass before a
galaxy reaches its peak gas mass is steeper than the decrease afterwards. The peak in the total cold gas content
coincides with a flattening in the stellar buildup of star-forming galaxies. The redshift at which the stellar mass dominates of the gas
mass in a galaxies decreases with decreasing $z=0$ host halo mass.

We clearly see that the evolution in the \hi and \h2 content of
galaxies changes with $z=0$ host halo mass as well. The cold gas is always
dominated by atomic hydrogen in galaxies
residing in haloes with $M_{\rm
  vir}\,_{z=0} \leq 10^{12}\,\rm{M}_\odot$. The dominance of \hi decreases as a
function of $z=0$ host halo mass. For galaxies residing more massive $z=0$ haloes the cold gas is first mostly molecular
and only at later times does the atomic hydrogen dominate. The
transition from an atomic to a molecular dominated interstellar medium
occurs at $z\sim 1$--$1.5$, after the peak in cold gas content of
galaxies. Figure \ref{fig:halo_evol} also shows that the evolution in
\hi mass at $z<1.5$ of galaxies is very minor, whereas the \h2 mass
decreases by at least an order of magnitude. Independent of halo mass
the fraction of baryons {\emph not} stored in the central galaxy
increases rapidly once the host halo reached a mass of $\sim 10^{12}\,\rm{M}_\odot$.

\section{Discussion}
\label{sec:discussion}
We have combined an abundance modeling approach \citep{Behroozi2013} with a method to
indirectly estimate galaxy gas masses \citep{Popping2012} to infer the
cold gas content of galaxies as a function of time and halo mass. It
has the unique capacity to obtain data-driven 
inferences of the \hi and \h2 mass of galaxies and
couple these to galaxy host halo mass.

\subsection{Mild evolution in the \hi mass of galaxies}
Our results suggest mild evolution in the \hi content of galaxies at fixed
stellar mass, whereas the \h2 content decreases rapidly. This picture
is supported by the evolution of the \hi and \h2 mass function. The
\h2 mass function shows a strong evolution of more than a dex with time
(especially in the redshift range $z=1-0$), whereas the \hi mass
function is fairly constant at $z<2$ (although note the drop in
  the normalization of the \hi
mass function at $z<0.5$ for galaxies with \hi masses larger than
$10^{10}\,\rm{M}_\odot$). Furthermore, we find very minor evolution
in the \hi mass of galaxies at $z<1.5$ as a function of their $z=0$
host halo mass. Within our model, the \h2
content of a galaxy is, to first order, controlled by 
the SFR (Equation \ref{eq:bigiel}). The \hi content of a galaxy is controlled by the amount of
cold gas available, combined with its partitioning into atoms and
molecules. This partitioning is set by the amount of gas
necessary to provide the pressure to support the formation of \h2 out
of which stars may form. The relatively weak evolution in galaxy \hi
content and the \hi mass function is the result of a self-regulated equilibrium that arises
naturally in galaxies. This equilibrium marks the balance between the
available cold gas content and the transformation of \hi into
\h2. Observations have shown that atomic hydrogen saturates at
surface densities of $\Sigma_{\rm HI} = 6 - 10\,\rm{M}_\odot\,\rm{pc}^{-2}$
and that cold gas at higher densities is dominated by molecular hydrogen
\citep{Blitz2006,Leroy2008}. As the cold gas content and surface
density of galaxies decreases due to gas consumption and disc growth, the amount of cold gas that is 
locked up in molecular hydrogen also rapidly decreases. This process may
prevent a rapid drop in the \hi reservoir of galaxies but does not
imply the weak evolution we find. It is the rate at
which galaxies loose their cold gas (through SF and/or feedback) and
their discs grow that sets the \h2 fraction of the cold gas and the weak evolution in galaxy \hi content. This provides an
interesting constraint for theoretical models that include the
partitioning of cold gas into \hi and \h2. 

\citet{Popping2013} found a similar evolution in the
  normalization and slope of the \hi mass function at $z < 2$. They  used various \h2
formation recipes in their semi-analytical model, among which the same
as employed in this work. 
Popping et al. also overpredict the number density
of galaxies with low \hi masses ($M_{HI} < 10^9\,\rm{M}_\odot$),
although not as dramatically as in Figure \ref{fig:massfunc_z0} and ascribe the excess of galaxies with low \hi masses to the inefficient conversion of atomic into molecular
hydrogen in low mass haloes ($M_{\rm vir} <
10^{10}\,\rm{M}_\odot$). Other semi-analytic approaches reached
similar conclusions \citep{Obreschkow2009_sam,Lagos2011cosmic_evol,Fu2012}.
Using a hydrodynamic code, \citet{Dave2013} found a strong evolution in the
slope of the low-mass end of the \hi mass function, unlike the results
presented in this work.  Dav\'e et al. use the same recipe to separate
 the atomic from the molecular gas as we us here. The authors find
 that the evolution in the low-mass-end of the \hi mass function
 directly follows the evolution of the stellar mass function, due to a
 constant ratio between the \hi and the stellar mass of galaxies.

\subsection{Galaxy gas fractions}
The cold gas fraction of a galaxy is set by the competition between the
accretion, heating, outflow, and the consumption of cold gas by star formation
\citep{Tinsley1980,Dave2011}. The evolutionary trend of cold gas fractions with
time therefore provides a unique insight into the importance of each
of these processes. We find that the cold gas fractions of galaxies
decrease at $z<3$ with the most rapid decrease found at $z<1.5$. This suggests that
the accretion of gas onto galaxies cannot cope with the
consumption and/or outflow of gas. Cosmological hydrodynamical
simulations predict a decline in their `cold mode' accretion rates onto galaxies
at $z<2$ \citep{Birnboim2003,Keres2005,Keres2009,Dekel2006,Dekel2009}. The
decrease in cold gas fraction at $z<1.5$ is in good agreement with a 
lack of accretion of new cold material onto galaxies. This supports a
picture in which below $z=2$ the infalling gas is sparse enough to
be heated up or blown away by shock heating or feedback from an active
galactic nucleus \citep[though see][]{Oppenheimer2010}.

We have seen that \frach2star evolves much more quickly than the
combined \hi$+$\h2 fraction of galaxies. The evolution of the atomic and
molecular gas mass of a galaxy do not necessarily go
hand-in-hand. The process that suppresses the formation of stars is at least two-fold: galaxies have lower
cold gas fractions and lower molecular fractions. This was
also found in other efforts to indirectly infer the gas mass of
galaxies \citep[Popping et al. in prep.]{Popping2012}. Previous
efforts were always based on a biased sample, whereas our
semi-empirical approach is unbiased. This conclusion is
supported by the decline in SFE with time at fixed stellar mass. Not
only the available gas reservoirs decrease, but the ability of the cold
gas to form stars also decreases.

We have found that galaxies with the highest stellar masses run out
of cold gas well before the lowest mass galaxies. This behavior is
also prominent in the evolution of \frach2star with stellar mass and
time. \citet{Popping2012} and Popping et al. (in 
prep.) found the same trend when looking at the indirectly estimated
gas content of galaxies taken from the COSMOS \citep{Scoville2007} and
CANDELS \citep{Grogin2011,Koekemoer2011} surveys. \citet{Popping2012} referred to this trend as a different manifestation of
downsizing in SF, which can be described as the decline in mass for
star-forming galaxies with increasing time \citep{Fontanot2009}. This behavior is supported by the evolution of the \h2
mass function: the number density of galaxies with high \h2 masses decreases
at earlier times than the number density of galaxies with low \h2
masses. This suggests that downsizing may be linked to the evolution of the
molecular gas content of galaxies. 

Theoretical models predict a similar evolution
for $f_{\rm gas}$ and \frach2star, although the normalization of the gas fraction
with time is lower. Models are known to form their stars
too early (especially in galaxies with low stellar mass, Weinmann et
al. 2012) and the low gas fractions may be a
reflection of the same physical process that drives the low-mass
galaxy problem \citep{White2014}. Our results have
the potential to act as a constraint on the physical recipes included
in models acting on the cold gas content of
galaxies. \citet{White2014} present the ``parking-lot'' picture as a
potential solution to the low-mass galaxy problem. In this picture,
when a halo accretes a smaller halo, the reservoir of ejected gas from
the minor progenitor is added to the reservoir of the ejected gas of
the new resulting halo (rather than adding it to the hot gas
reservoir as was previously done in semi-analytic models). The model
then adopts various scaling relations for the time after which the gas parked in this reservoir of ejected gas can accrete into
the halo and cool onto the galaxy. This gas cools down and participates in the
formation of stars at a later time when adopting the
``parking-lot'' picture than in the previous versions of the
semi-analytic model. \citet{White2014} showed that this process
increases the predicted gas fractions of galaxies at our redshifts of
interest, in better agreement with our inferred gas fractions.

\subsection{Cosmic densities}
We obtain good agreement with the observed values for the cosmic \hi density
at $z\approx 2$--$3$ and correctly reproduce the observational constraints
on the cosmic \h2 density at redshifts $z=1.5$ and $z=2.75$. The
relatively constant evolution in $\Omega_{\rm HI}$ at $z< 2$ is in sharp
contrast with the evolution of the cosmic star-formation rate,
whereas the molecular hydrogen density (by construction) tightly
follows the SFR density and decreases by over an order of magnitude in
the same redshift regime. This suggests that in a model where the
SFR of a galaxy is controlled by the available molecular gas, one can
simultaneously reproduce the peak in cosmic SFR and the weak evolution of the \hi density of the Universe during the
same cosmic epoch. A similar conclusions was reached using theoretical
models \citep{Obreschkow2009_sam,Bauermeister2010,Lagos2011cosmic_evol,Popping2013},
but this has not been addressed before through a data-driven approach as
presented in this work.

\subsection{On the gas properties that drive the peak in the stellar
  mass - halo mass ratio}
Abundance matching approaches have demonstrated the importance of the
galaxy-halo connection. Haloes near $10^{12}\,\rm{M}_\odot$ are most
efficient turning their baryons into stars \citep{Behroozi2013SFE}. This maximum efficiency is supported by the gas content of
galaxies. Haloes with masses near the characteristic mass of
$10^{12}\,\rm{M}_\odot$ have the highest gas-to-halo mass ratio. The
\h2-to-halo mass ratio shows an especially strong
peak near $10^{12}\,\rm{M}_\odot$. This is not necessarily
due to small cold gas reservoirs at other halo masses. Galaxies residing in lower mass
haloes are dominated by their cold gas content, but have very low molecular fractions. As a
result a relatively small amount of gas is available for the formation of
stars. This suggests an interesting physical scenario, in which these
galaxies are efficient at cooling gas onto their disc, but physics
internal to the galaxies prevents them from forming molecules. That
is, low densities, stellar winds and supernova feedback seem to be efficient at  disrupting \h2 formation, and are not (nor do they need to be) efficient at suppressing cooling for HI even at $z=0$.

We have shown that galaxies residing in haloes larger than $10^{12}\,\rm{M}_\odot$ cannot
build up gas reservoirs much larger than a few times
$10^{10}\,\rm{M}_\odot$. We also see this in the time
evolution of the gas content in galaxies (Figure
\ref{fig:halo_evol}). This is
supported by the decline in cold gas fractions as a function of halo
mass seen at all redshifts. Interestingly, the cold gas molecular fraction is roughly constant for haloes with 
masses larger than $10^{12}\,\rm{M}_\odot$ as is the star formation
efficiency of the galaxies. This constancy in \h2 fraction is supported by the drop in cold gas fraction and \frach2star for
galaxies with masses $M_{\star} > 10^{10}\,\rm{M}_\odot$. This upper
limit in cold gas molecular fraction provides an explanation for the
plateau in SFE for galaxies with stellar masses $M_{\star} >
10^{10}\,\rm{M}_\odot$. 

Haloes of masses $\sim 10^{12}\,\rm{M}_\odot$ have stored approximately 20--30
per cent of their baryons in stars and cold gas in the central galaxy. Haloes with lower and
higher masses are less efficient in bringing the baryons down to their
central galaxy. This supports the findings that
star-forming and passive galaxies have significant ionized gas
reservoirs in their haloes \citep{Tumlinson2011,Bordoloi2014,Werk2014}. The masses of
these cold gas reservoirs are comparable for star-forming and passive
galaxies \citep{Thom2012}. All of these results tie in with a picture in which heating processes
(such as supernovae feedback, active galactic nuclei feedback, and shock
heating) prevent the build-up of large amounts of cold gas onto the
galaxies. Especially once a host halo reaches a mass of $\sim 10^{12}
\,\rm{M}_\odot$ the number of baryons not stored in the central
galaxy grows rapidly (Figure \ref{fig:halo_evol}). However, once cooled onto a galaxy the local gas conditions
determine the efficiency of the cold gas to become molecular and form
stars.

We note that our present conclusions are based on our assumptions that
  \h2 traces SF and that the partitioning of cold gas in \hi and
\h2 only depends on the pressure acting on the gas. Processes such
as ram-pressure stripping and the destruction of molecules by AGN feedback
also affect the gaseous and molecular content of
galaxies. Furthermore, dust grains and metals play an important role
as the catalysts and coolants for \h2 formation. This is especially important in low-density and low-metallicity environments \citep{Krumholz2011,Popping2013} and
would further suppress the formation of molecules in the lowest mass haloes.

\subsection{Excess of galaxies with low \hi masses}
\label{sec:HI_excess}
Our model correctly reproduces the \hi and \h2 content of local
galaxies and the observed \h2 mass function. We obtain good agreement
with the observed \hi mass function for \hi-rich galaxies ($M_{\rm HI} > 10^{9.0}\,\rm{M}_\odot$), but
overpredict the number of galaxies with lower \hi masses. The galaxies
responsible for this excess have low stellar masses ($M_{\star} <
10^{8.5}\,\rm{M}_\odot$). To first order the steep slope
at the low-mass end of the \hi mass function is a result of the steep slope in the
predicted stellar mass function \citep{Behroozi2013}, corresponding
with an upturn in the observed stellar mass function below $10^{8.5}
\rm{M}_\odot$ \citep{Baldry2008}. The exact shape of the stellar mass
function in this regime is quite uncertain due to incompleteness of
low-surface-brightness galaxies \citep{Baldry2012}. 

{Nevertheless, it is
highly uncertain whether the poor number statistics of low-surface-brightness
galaxies can fully account for the severe mismatch between the observed and
inferred \hi mass function. For galaxies with low stellar masses
  the observed ratio between the \hi and stellar mass of galaxies has
  a steep faint-end slope \citep{McGaugh2012,Zhang2012}. The stellar mass
function has a negative slope with stellar mass in the same mass
regime \citep{Baldry2008,Baldry2012}. It is then hard to reconcile
that the observed \hi mass function is relatively flat at
\hi masses of $10^{8-9}\,\rm{M}_\odot$ \citep{Zwaan2005,Martin2010}. We
can only explain for this if the \hi-to-stellar mass ratio
for galaxies with low stellar masses is not as well
behaved as suggested by current surveys and we are
missing a significant population of \hi poor galaxies at low stellar
masses (for example dwarf spheroidal galaxies). Given the steep slope we find in the \h2 mass function at
$M_{\rm \star} < 10^8\,\rm{M}_\odot$ it is not unlikely that these \hi
poor galaxies are also diluted from \h2.

We note that our assumption for exponential discs may
not be appropriate for galaxies with \hi masses smaller than
$10^9\,\rm{M}_\odot$. \citet{Kelvin2014} showed that below
stellar masses of $10^{9}\,\rm{M}_\odot$ the stellar mass
function is mostly made up by irregular galaxies. The different
distribution of gas in these galaxies compared to our assumption of
exponential discs can have a significant effect on the inferred \hi
masses. Furthermore, the stellar and gas sizes of these low mass galaxies are
  observationally not well
constrained. A better understanding of the mass and distribution of \hi for
a large sample of galaxies with low stellar masses can improve the
situation and resolve the discrepancy between the observed and
inferred \hi mass function at $z=0$.

\section{Summary \& Conclusions}
\label{sec:summary}
In this paper we have presented a new semi-empirical approach to
infer the \hi and \h2 content of galaxies as a function of redshift and
host halo mass. All data results in this paper are available for
download
online.\footnote{\url{http://www.eso.org/~gpopping/Gergo_Poppings_Homepage/Data.html}}
We summarize our main results below:

\begin{itemize}
\item There is weak evolution in the \hi content and \hi mass function of galaxies
  at redshifts $z<1.5$ (less than half a dex in \hi mass), whereas the \h2 content and mass function evolve
  strongly in this redshift range (more than an order of magnitude in
  \h2 mass at $z < 1.5$). This behavior originates in a
  self-regulated equilibrium in the \hi mass of galaxies controlled
  by the amount of cold gas available and its partitioning into an
  atomic and molecular part.

\item The cold gas fraction of galaxies decreases as a function of
  stellar mass and time. The relative \h2 content of galaxies
  decreases with stellar mass and redshift as well, but more
  rapidly (from 0.75 to almost zero at stellar masses of
  $10^{10}\,\rm{M_\odot}$). 

\item Massive galaxies consume and/or expel their cold gas content
  earlier than less massive galaxies. Galaxies with a stellar mass of
  $10^{11}\,\rm{M_\odot}$ are dominated by their stellar content at
  $z=3$, whereas galaxies with a stellar mass of $10^{9}\,\rm{M_\odot}$
  still have a gas fraction of 50\% at $z=0$. We find a rapid decrease in cold
  gas fractions at $z<1.5$. This supports a picture in which the
  accretion of new cold material onto galaxies slows down at these
  redshifts, possibly due to shock heating or heating from active
  galactic nuclei feedback of the sparse infalling gas.

\item The SFR-to-gas mass ratio, i.e. the star formation efficiency, decreases by more than an order of
  magnitude from $z=3$ to $z=0$. This is driven by the declining trend
  of molecular fractions with time, allowing relatively less gas to form stars.

\item We confirm that when adopting a \h2-based star-formation
  relation one can fuel the peak in cosmic SFR at
  $z\sim 2$ and simultaneously reproduce the weak evolution in $\Omega_{\rm HI}$
  during the same epoch.

\item Galaxies residing in haloes of mass $\sim 10^{12}\,\rm{M}_\odot$ are
  most efficient in attaining cold gas \emph{and} forming molecules. The
  combination of these two effects explains for the high star-formation
  efficiencies of galaxies in haloes in the same mass range.  In
  haloes of mass $\sim 10^{12}\,\rm{M}_\odot$, roughly 30 percent of the baryons is locked up in cold gas and
  stars within the central galaxy at $z=3$ . This number decreases to roughly 17 percent for galaxies
  residing in haloes with the same mass at $z=0$. Galaxies in
  lower-mass haloes have significant cold gas reservoirs, but only a
  minor fraction of their gas is molecular. Heating processes prevent
  the buildup of large gas reservoirs in galaxies residing in more
  massive haloes.
\end{itemize}

These results serve as predictions for future surveys of the atomic and molecular content of galaxies. We
look forward to observations from new and upcoming radio and sub-mm
facilities such as ALMA, SKA and its pathfinders MeerKat and ASKAP
that will confront our results. The inferred gas masses presented do
not depend on unknown recipes for feedback and heating processes,  but
on galaxy properties that by construction are representative of real
galaxies. We anticipate that semi-empirical efforts such as presented
here have the potential to break degeneracies between different
physical recipes, allowing for a better  understanding of the fueling
and stellar-buildup of galaxies. 

\section*{Acknowledgments}
We thank the anonymous referee for constructive comments. We
thank Romeel Dav\'e, Thijs van der Hulst, Andrey Kravtsov, Houjun Mo, Rachel
Somerville, Marco Spaans, Scott Trager, Arjen van der Wel, and Cathy White for
stimulating discussions and comments, and Amber Bauermeister, Matthew
Bothwell, Claudia Lagos, Danail Obreschkow, and Fabian Walter for providing data. GP acknowledges NOVA (Nederlandse Onderzoekschool voor Astronomie) for funding. PB was supported by a
Giacconi Fellowship through the Space Telescope Science Institute,
which is operated by the Association of Universities for Research in
Astronomy, Incorporated, under NASA contract NAS5-26555. We also thank galaxies
for being awesome.
\bibliographystyle{mn2e_fix}
\bibliography{references}

\appendix
\section[]{Gas evolution for individual galaxies}
We presented the evolution of the cold gas content of galaxies as a
function of their $z=0$ host halo mass in Figure
\ref{fig:halo_evol}. The presented results are based on median
accretion histories, smoothing over the accretion and star formation
histories of individual galaxies. We present the evolution of cold gas
content for individual mass accretion histories in
Figure \ref{fig:halo_evol_ind}. The trends of cold gas, \hi, and \h2
with time are similar to the trends showed in Figure
\ref{fig:halo_evol} but with a wider spread.
\begin{figure*}
\includegraphics[width = 0.9\hsize]{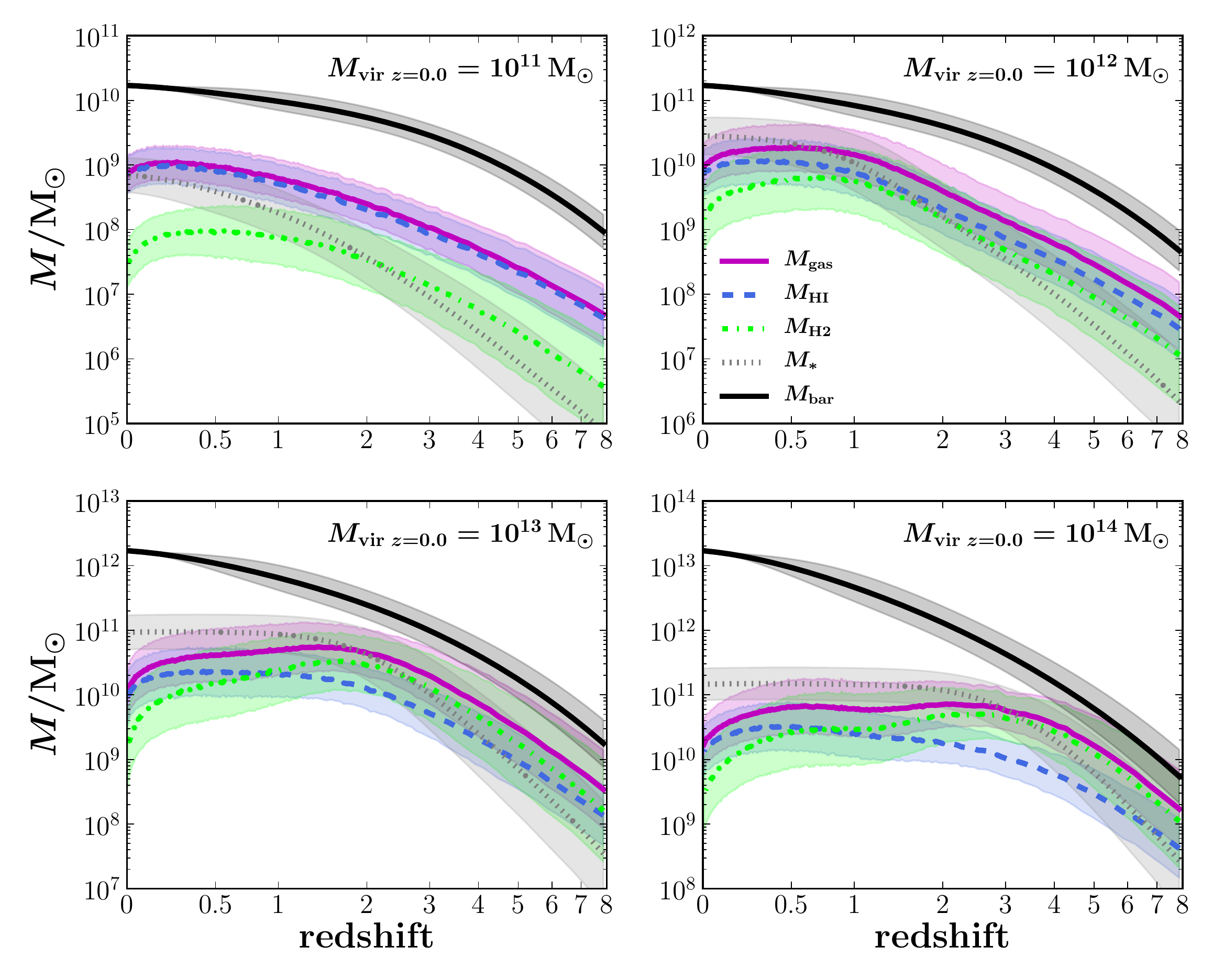}
\caption{Time evolution of the total cold gas (solid purple), \hi
  (dashed blue), \h2 (dashed dotted green), stellar
  (dashed grey) and baryonic mass of the halo (solid black) for haloes as a
  function of their $z=0$ halo mass for individual mass accretion
  histories (rather than the median). The shaded regions show the one-sigma posterior
  distributions.\label{fig:halo_evol_ind}}
\end{figure*}

\end{document}